\documentstyle[preprint,tighten,eqsecnum,floats,aps,epsfig]{revtex}

\begin{document}

\draft

\title{ Antisymmetrized random phase approximation for quasielastic \\
 scattering in nuclear matter: Non-relativistic potentials
}
\author{
A. De Pace 
}
\address{
 Istituto Nazionale di Fisica Nucleare, Sezione di Torino, \\
 via P. Giuria 1, I-10125 Torino, Italy
}
\date{February 1998}

\maketitle

\begin{abstract}
Many-body techniques for the calculation of quasielastic nuclear matter response
functions in the fully antisymmetrized random phase approximation on a
Hartree-Fock basis are discussed in detail. The methods presented here allow for
an accurate evaluation of the response functions with little numerical effort.
Formulae are given for a generic non-relativistic potential parameterized in
terms of meson exchanges; on the other hand, relativistic kinematical effects
have been accounted for.
\end{abstract}
\pacs{PACS: 21.60.Jz, 21.65.+f, 24.10.Cn \\
Keywords: Hartree-Fock and random phase approximation; Nuclear matter;
Quasielastic scattering }

\section{ Introduction }
\label{sec:intro}

Quasielastic electron scattering on nuclei has been in the past years the 
subject of intense experimental \cite{Jou96,Ang96,Wil97} and theoretical 
(see, e.~g., Refs.~\cite{Del85,Del87,Shi89,Hor90,Bub91,Weh93,Ama94,Cai95,
Kim95,Bar96a,Bes96,Fab97,Cen97,Gil97}) investigations. The first aim of the
theoretical studies is to test the available nuclear models; once the nuclear
physics issues are well understood, one might hope to gain insight into other 
aspects of the problem, for instance by extracting with sufficient precision the
nucleon form factors.

In principle, the quasifree regime makes one confident that the physical
quantities of interest may be computed in a reliable way; in practice, also in
this case one has to cope with considerable computational problems.
Many diverse techniques have been employed in the literature. Each of them has
its own relative merits and deficiencies and, in general, it would be highly
desirable to be able to reach some degree of convergence in their outcomes.

In the following, we shall be concerned with Green's function techniques, as
introduced, e.~g., in Ref.~\cite{Fet71}. These methods can be, and have been,
applied both to finite nuclei and nuclear matter, the choice being generally
driven by the specific reaction and by the momentum regime of interest.
Here, we shall focus on the nuclear matter, having in mind applications to
electron scattering (that is, without the complications introduced by the
reaction mechanism of hadronic probes) in a range from a few hundreds to several
hundreds MeV/c of transferred momenta (where the quasielastic peak is
sufficiently far from low-energy resonances and not too much affected by finite 
size effects). The use of nuclear matter reduces the computational load, thus
allowing a more straightforward implementation of more sophisticated theoretical
schemes: This makes easier to develop and test approximation methods that could
then be utilized also for calculations in finite nuclei.

Let us now briefly browse the theoretical framework that we shall discuss in
detail in the following sections.

A first choice one has to do in setting up the formalism concerns the treatment
of relativistic effects. Trivial kinematical effects can be obviously rather
important and can be included in a straightforward way.
The treatment of dynamical effects is more delicate. Two main paths have been
followed in the literature: Either using field theoretical methods (as done, 
e.~g., in the Walecka model and its derivations \cite{Wal95}) or potential
techniques (using, i.~e., phenomenological potentials truncated at some order in
the non-relativistic expansion). Here, we shall put ourselves on the second
path, but, to contain the amount of material, we shall employ strictly
non-relativistic potentials. The extensions necessary to include higher order
relativistic terms will be discussed elsewhere (see, however, 
Refs.~\cite{Bar96a,Bar96b,Amo96} for a few applications).

Next, one should choose the phenomenological input potential and, in connection
with this choice, possibly the way of dealing with short-range correlations.
All the formulae we are going to give in the following sections are based on a
generic one-boson-exchange potential. They can thus be used both with a bare
phenomenological interaction, --- such as one of the Bonn potential variants,
--- or with a one-boson-exchange parameterization of a $G$-matrix generated from
some potential. The use of an effective interaction derived from a $G$-matrix is
a common way of including short-range correlations. One should be aware of
possible problems due to the use of a local potential to fit non-local matrix
elements. At least in a few cases discussed in the literature this does not
appear to be a reason of concern \cite{Nak84,Nak87}.
On the other hand, possible effects due to the specific quasielastic regime
remain completely unexplored: Indeed, $G$-matrices employed in quasielastic
calculations are usually generated using bound state boundary conditions, which
make them real and practically energy independent, while, in general, they might
be complex and energy dependent.

Once we have fixed the effective interaction, we can proceed to consider a 
hierarchy of approximation schemes.

The lowest order approximation is, of course, given by the free Fermi gas.
Then, one may include mean field correlations at the Hartree-Fock (HF) level
(or Brueckner-Hartree-Fock (BHF) if short-range correlations are accounted for).
In nuclear matter a HF calculation can be done exactly without too many efforts.
Nevertheless, we show how a quite accurate analytic approximation can be
derived, since we shall need the method later to combine the HF and the random
phase approximation (RPA) schemes.
The latter is the last resummation technique to be discussed. 
It should be noticed that even in nuclear matter the calculation of the {\em
antisymmetrized} RPA response functions is not trivial.
Indeed, most calculations that are labeled ``RPA'' in the literature are
actually performed in the so-called ``ring approximation'', where only the
direct contributions are kept: In this case, in nuclear matter one gets
an algebraic equation. Here, we use the continued fraction (CF) technique to 
provide a semi-analytical estimate of the full RPA response (see 
Refs.~\cite{Shi89} and \cite{Bub91} for alternative methods).
Calculations with this method have been performed both in finite nuclei 
\cite{Del85,Del87} and in nuclear matter \cite{Alb93,Bar94,Bar96a,Bar96b}, 
always truncating the CF expansion at first order, because of the difficulty of 
the numerical calculations. We have pushed the analytical calculation far enough
to allow not only a fast and accurate estimate of the first order CF expansion, 
but also of the second order one. Since in the CF technique there is no general 
way of estimating the convergence of the series, this is the only way of getting
a quantitative hold on the quality of the approximation.
As noticed before, HF (and kinematical relativistic) effects can then be 
incorporated in the RPA calculation, yielding as the final approximation scheme 
a HF-RPA (or BHF-RPA) response function.

Of course, many diverse many-body contributions have been left out.
It should however be noted that the classes of many-body diagrams discussed
here, on the one hand already allow one to study many interesting features of
the quasielastic response; on the other hand, the fact of having developed
semi-analytical methods reduces to a minimum the computational efforts, thus
making this formalism a good starting point for the study of other many-body
effects.

The paper is organized as follows. In Section 2 the theoretical machinery is set
up, discussing in separate subsections the treatment of relativistic kinematics
and the free, HF and RPA responses. The intent of this paper is just to provide
theoretical tools, so we do not attempt any discussion of the phenomenology of
quasielastic scattering. On the other hand, in Section 3 calculations based on 
the formalism previously developed are shown, in order to test the convergence 
of the CF expansion and the importance of antisymmetrization.
Finally, in the last Section we present a few concluding remarks.

\section{ Response functions }
\label{sec:Resp}

Let us consider an infinite system of (possibly) interacting nucleons, at some
density corresponding to a Fermi momentum $k_F$. For the kinetic energies of
the nucleons we can choose either the relativistic or non-relativistic
expressions, whereas we assume that the interactions take place through a
non-relativistic potential. For the latter we take the following general form
in momentum space
\begin{eqnarray}
  V(\bbox{k}) &=& V_0(k) +
    V_{\tau}(k) \bbox{\tau}_1\cdot\bbox{\tau}_2 +
    V_{\sigma}(k) \bbox{\sigma}_1\cdot\bbox{\sigma}_2 +
    V_{\sigma\tau}(k) \bbox{\sigma}_1\cdot\bbox{\sigma}_2 \ 
      \bbox{\tau}_1\cdot\bbox{\tau}_2 \nonumber \\
  && \quad +
    V_t(k) S_{12}(\hat{\bbox{k}}) +
    V_{t\tau}(k) S_{12}(\hat{\bbox{k}}) \bbox{\tau}_1\cdot\bbox{\tau}_2 ,
\label{eq:pot}
\end{eqnarray}
where $S_{12}$ is the standard tensor operator and $V_{\alpha}(k)$ represents
the momentum space potential in channel $\alpha$.
Here, we assume that $V_{\alpha}(k)$ has the general form of a static
one-boson-exchange potential, so that in each spin-isospin channel it is given
as a sum of contributions from different mesons, 
$V_{\alpha}\equiv\sum_i V_{\alpha}^{(i)}$. In the central channels ($0$, $\tau$,
$\sigma$, $\sigma\tau$) the contribution from any meson can be expressed as the
combination of a short-range (``$\delta$'') piece and a longer range (``momentum
dependent'') piece\footnote{ The nomenclature stems from the fact that, in the
absence of form factors, $V_{\delta}$ is a constant and is represented by a
Dirac $\delta$-function in coordinate space, whereas $V_{\text{MD}}$ is,
indeed, the momentum dependent piece.}:
\begin{mathletters}
\label{eq:mes-exch}
\begin{eqnarray}
  V_{\delta}^{(i)}(k) &=&  g_{\delta}^{(i)} 
    \left(\frac{\Lambda_i^2-m_i^2}{\Lambda_i^2+k^2}\right)^\ell \\
  V_{\text{MD}}^{(i)}(k) &=&  g_{\text{MD}}^{(i)} 
    \frac{m_i^2}{m_i^2+k^2}
    \left(\frac{\Lambda_i^2-m_i^2}{\Lambda_i^2+k^2}\right)^\ell , 
    \quad  \ell=0,1,2 ,
\end{eqnarray}
whereas in the tensor channels ($t$, $t\tau$) is given by
\begin{equation}
  V_{\text{TN}}^{(i)}(k) =  g_{\text{TN}}^{(i)} 
    \frac{k^2}{m_i^2+k^2}
    \left(\frac{\Lambda_i^2-m_i^2}{\Lambda_i^2+k^2}\right)^\ell , 
    \quad \ell=0,1,2 .
\end{equation}
\end{mathletters}
In Eqs.~(\ref{eq:mes-exch}), $g_{\delta}^{(i)}$, $g_{\text{MD}}^{(i)}$ and 
$g_{\text{TN}}^{(i)}$ are the (dimensional) coupling constant of the $i$-th
meson, $m_i$ is its mass and $\Lambda_i$ the cut-off; to be more general, we
have allowed for a choice among potentials without form factors or with monopole
or dipole form factors.

\begin{figure}[t]
\begin{center}
\mbox{\epsfig{file=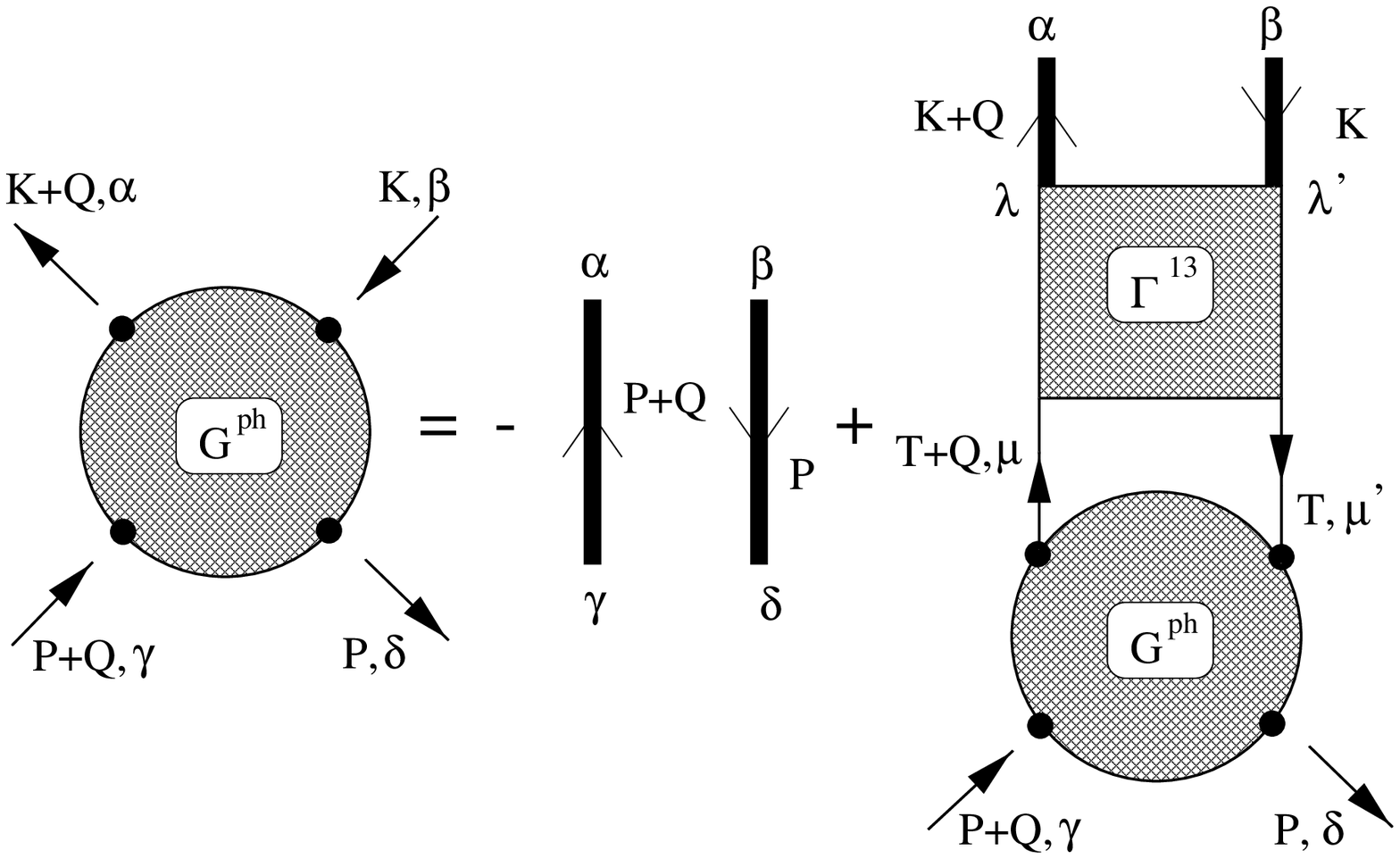,width=.7\textwidth}}
\vskip 2mm
\caption{ Diagrammatic representation of the Galitskii-Migdal integral equation
for the ph Green's function, $G^{\text{ph}}$; $\Gamma^{13}$ is the irreducible
vertex function in the ph channel; the heavy lines represent the exact one-body
Green's functions.
 }
\label{fig:Gph}
\end{center}
\end{figure}

Our starting point \cite{Gal58,Abr63,Alb91} is given by the Galitskii-Migdal 
integral equation for the particle-hole (ph) four-point Green's 
function\footnote{ Capital letters refer to four-vectors; small case letters 
to three-vectors; the Greek letters $\alpha,\beta,...$ refer to a set of 
spin-isospin quantum numbers.},
\begin{eqnarray}
  && G^{\text{ph}}_{\alpha\beta,\gamma\delta}(K+Q,K;P+Q,P) = 
    -G_{\alpha\gamma}(P+Q)\,G_{\delta\beta}(P)\,(2\pi)^4\delta(K-P) \nonumber \\
  && + i G_{\alpha\lambda}(K+Q)\,G_{\lambda'\beta}(K)\int\frac{d^{4}T}{(2\pi)^4}
    \Gamma^{13}_{\lambda\lambda',\mu\mu'}(K+Q,K;T+Q,T)\,
    G^{\text{ph}}_{\mu\mu',\gamma\delta}(T+Q,T;P+Q,P) , \nonumber \\
\label{eq:Gph}
\end{eqnarray}
which is diagrammatically illustrated in Fig.~\ref{fig:Gph}.
In (\ref{eq:Gph}), $G$ represents the exact one-body Green's function, whereas 
$\Gamma^{13}$ is the irreducible vertex function in the ph channel.

\begin{figure}[t]
\begin{center}
\mbox{\epsfig{file=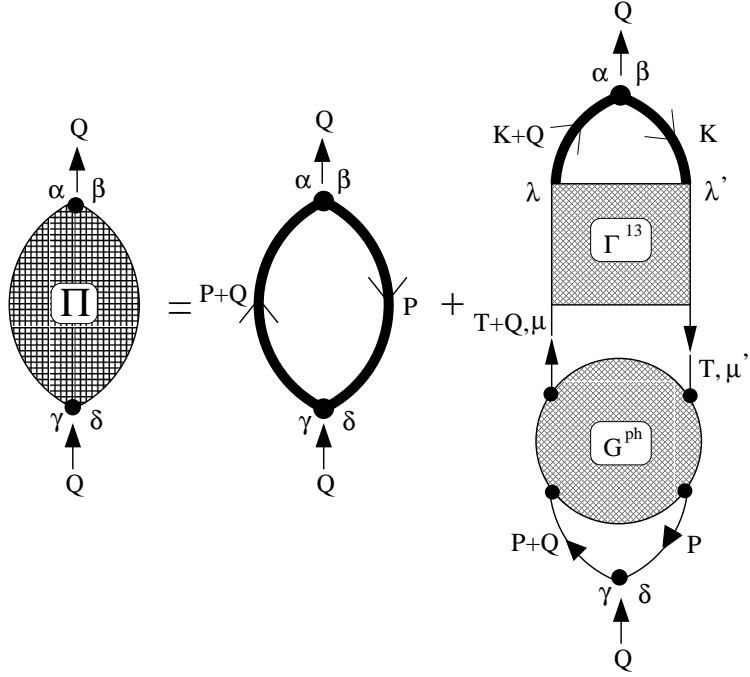,width=.6\textwidth}}
\vskip 2mm
\caption{ Diagrammatic representation of the polarization propagator $\Pi$
derived from the ph Green's function $G^{\text{ph}}$.
 }
\label{fig:Pi}
\end{center}
\end{figure}

Given $G^{\text{ph}}$ one can then define the {\em polarization propagator}
\begin{eqnarray}
  \Pi_{\alpha\beta,\gamma\delta}(Q) & \equiv & 
    \Pi_{\alpha\beta,\gamma\delta}(q,\omega) \nonumber \\
  &=& i\int\frac{d^{4}P}{(2\pi)^4}\frac{d^{4}K}{(2\pi)^4}
    G^{\text{ph}}_{\alpha\beta,\gamma\delta}(K+Q,K;P+Q,P) ,
\label{eq:Pi}
\end{eqnarray}
whose diagrammatic representation is displayed in Fig.~\ref{fig:Pi}.
Note that for $\Pi(q,\omega)$ one cannot, in general, write down an integral (or
algebraic) equation.

In the case of electron scattering, one can define charge, --- or {\em
longitudinal}, --- and magnetic, --- or {\em transverse}, --- polarization
propagators:
\begin{mathletters}
\label{eq:PiLT}
\begin{eqnarray}
  &&\Pi^{I}_{\text{L}}(q,\omega) = \text{tr}[\hat{O}^{I}_{\text{L}}
    \hat{\Pi}(q,\omega) \hat{O}^{I}_{\text{L}}] \\
  &&\Pi^{I}_{\text{T}}(q,\omega) = \sum_{ij}\Lambda_{ji}\Pi^{I}_{ij}(q,\omega)
    , \qquad \Pi^{I}_{ij}(q,\omega) = \text{tr}[\hat{O}^{I}_{\text{T};i}
    \hat{\Pi}(q,\omega) \hat{O}^{I}_{\text{T};j}] \\
  &&\phantom{\Pi^{I}_{\text{T}}(q,\omega) = 
    \sum_{ij}\Lambda_{ji}\Pi^{I}_{ij}(q,\omega) , } \qquad \,
    \Lambda_{ij} = (\delta_{ij}-\hat{\bbox{q}}_i\hat{\bbox{q}}_j)/2 , 
    \nonumber
\end{eqnarray}
\end{mathletters}
where, for brevity, the dependence upon the spin-isospin indices has been
represented in matrix form, introducing hats where appropriate.
In (\ref{eq:PiLT}), $I$ labels the isospin channel and the longitudinal and
transverse vertex operators are given as follows:
\begin{equation}
  \left\{ 
    \begin{array}{l}
      \hat{O}^{I=0}_{\text{L}} = 1/2 \\
      \hat{O}^{I=1}_{\text{L}} = \tau_3/2 
    \end{array}
    \right. \qquad\qquad
  \left\{
    \begin{array}{l}
      \hat{O}^{I=0}_{\text{T};i} = \sigma_i/2 \\
      \hat{O}^{I=1}_{\text{T};i} = \sigma_i\tau_3/2 .
    \end{array}
    \right.
\label{eq:OLT}
\end{equation}
The quantity of interest here is the imaginary part of 
$\Pi_{\text{L,T}}(q,\omega)$, since the inelastic scattering cross section,
--- where the momentum $q$ and the energy $\omega$ have been transferred to the
nucleus, --- is a linear combination of $\text{Im}\Pi_{\text{L}}$ and
$\text{Im}\Pi_{\text{T}}$. It is then customary to define longitudinal and
transverse response functions
\begin{equation}
  R_{\text{L,T}}(q,\omega) = R^{I=0}_{\text{L,T}}(q,\omega) + 
    R^{I=1}_{\text{L,T}}(q,\omega) ,
\end{equation}
which are related to $\Pi_{\text{L,T}}$ by
\begin{eqnarray}
  R^{I}_{\text{L,T}}(q,\omega) &=& 
    -\frac{V}{\pi} {f^{(I)}_{\text{L,T}}}^2(q,\omega)
    \text{Im}\Pi^{I}_{\text{L,T}}(q,\omega) \nonumber \\
  &=& -\frac{3\pi A}{2k_F^3} {f^{(I)}_{\text{L,T}}}^2(q,\omega) 
    \text{Im}\Pi^{I}_{\text{L,T}}(q,\omega) ,
\label{eq:RILT}
\end{eqnarray}
where $V$ is the volume, $A$ the mass number and ${f^{(I)}_{\text{L,T}}}^2$ 
contain the squared electromagnetic form factors of the nucleon.

\subsection{ Non-relativistic vs relativistic kinematics}
\label{subsec:rel}

The response functions introduced above have been defined as functions of the
momentum transfer $q$ and of the energy transfer $\omega$.
Actually, it is possible, --- and convenient, --- to define a scaling variable
$\psi$, which combines $q$ and $\omega$: This variable is such that the free
responses in the non-Pauli-blocked region ($q>2 k_F$) can be expressed in terms 
of the unique variable $\psi$ (apart from $q$-dependent multiplicative factors).
We shall see that even in the Pauli-blocked region and for an interacting system
it is convenient to use the pair of variables ($q$,$\psi$) instead of 
($q$,$\omega$).

Besides the obvious advantages related to the use of a scaling variable, there
is another good reason for expressing the responses in terms of $\psi$: In fact,
in this way one can define response functions that are independent of the form
chosen for the nucleon kinetic energy.
To be more specific: Starting from either a non-relativistic or a relativistic
Fermi gas, one is always lead to the same expressions for the responses in terms
of ($q$,$\psi$)\footnote{ Strictly speaking, the validity of this statement is
approximate, but quantitatively accurate.}; to be different in the two cases is,
of course, the definition of $\psi$ in terms of ($q$,$\omega$).

We shall see in the following subsections that the energy denominators of the
free nucleon propagators, appearing in the Feynman diagrams for the response
functions, can always be written as 
$\omega-\epsilon^{(0)}_{\bbox{k}+\bbox{q}}+\epsilon^{(0)}_{\bbox{k}}$, where 
$\epsilon^{(0)}_{\bbox{k}}$ is the kinetic energy of a nucleon of momentum $k$
and $k<k_F$. In the non-relativistic case, one finds
\begin{equation}
  \omega-\epsilon^{(0)\text{nr}}_{\bbox{k}+\bbox{q}}
    +\epsilon^{(0)\text{nr}}_{\bbox{k}} = 
  \frac{qk_F}{m_N}\left(\psi_{\text{nr}}-
    \hat{\bbox{q}}\cdot\frac{\bbox{k}}{k_F}\right) ,
\label{eq:endenNR}
\end{equation}
where
\begin{equation}
  \psi_{\text{nr}} = \frac{1}{k_F}\left(\frac{\omega m_N}{q}-\frac{q}{2}\right)
\label{eq:psiNR}
\end{equation}
is the standard scaling variable of the non-relativistic Fermi gas and $m_N$
the nucleon mass.

In the relativistic case, in Ref.~\cite{Alb90} it had been shown that at the 
pole it is a very good approximation to use Eq.~(\ref{eq:endenNR}) substituting 
$\psi_{\text{nr}}$ with 
\begin{equation}
  \psi_{\text{r}} = \frac{1}{k_F}\left[
    \frac{\omega m_N(1+\omega/2m_N)}{q}-\frac{q}{2}\right] 
\label{eq:psiR}
\end{equation}
and multiplying the free response by the Jacobian of the transformation,
$1+\omega/m_N$.

\begin{figure}[t]
\begin{center}
\vskip 2mm
\mbox{\epsfig{file=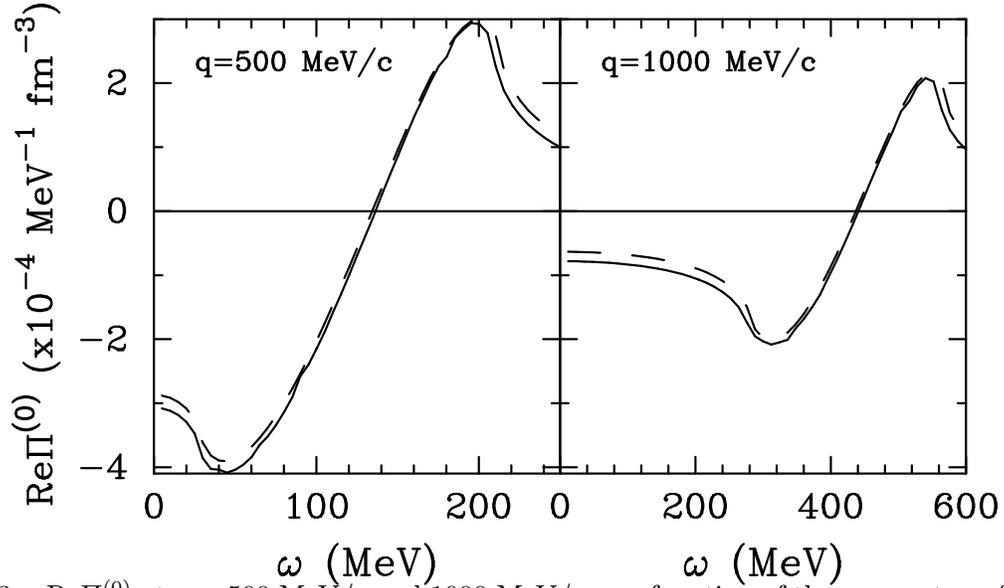,width=.8\textwidth}}
\caption{ $\text{Re}\Pi^{(0)}$ at $q=500$ MeV/c and 1000 MeV/c as a function of
the energy transfer: Using the exact relativistic kinetic energies (solid) and
the approximation discussed in the text (dash); $k_F=195$ MeV/c.
 }
\label{fig:RePi0}
\end{center}
\end{figure}

However, in the calculation of higher order (RPA) contributions, also the real
part of the energy denominators comes into play and one has to check the quality
of the approximation far from the pole. 
With some algebra, --- and assuming $k^2/m_N^2<<1$, --- one can write 
\begin{eqnarray}
  \omega-\epsilon^{(0)\text{r}}_{\bbox{k}+\bbox{q}}
    +\epsilon^{(0)\text{r}}_{\bbox{k}} &\cong& 
    \frac{q k_F}{m_N} 
    \frac{\psi_r-\hat{\bbox{q}}\cdot\bbox{k}/k_F}{\displaystyle 
      \frac{\strut 1}{\displaystyle 2}
      \left(1+\frac{\strut\omega}{\displaystyle m_N}
        +\sqrt{1+\frac{\strut q^2+2\bbox{q}\cdot{k}}{\displaystyle m_N^2}}
        \right)}
  \nonumber \\
  &\cong& \frac{q k_F}{m_N} 
    \frac{\psi_r-\hat{\bbox{q}}\cdot\bbox{k}/k_F}{1+\omega/m_N} ,
\label{eq:endenRr}
\end{eqnarray}
where, in the last passage, we have replaced the square root with its value at
the pole.
In Fig.~\ref{fig:RePi0}, we display the real part of the free polarization 
propagator (defined in the following subsection) using the exact relativistic 
dispersion relation and the prescription of Eq.~(\ref{eq:endenRr}) at $q=500$ 
MeV/c and 1 GeV/c as a function of $\omega$. We note that the agreement between
the two ways of calculating $\text{Re}\Pi^{(0)}$ is quite good at both momenta.

Eq.~(\ref{eq:endenRr}) provides an approximation for the free ph propagator: A
prescription to get the (kinematically) relativistic polarization propagators at
any order in the RPA expansion (see Section \ref{subsec:RPA-resp}) can easily be
obtained by noting that $\Pi^{(n)}$, --- the $n$-th order contribution to the
RPA chain, --- contains $n+1$ ph propagators; then, one has
\begin{mathletters}
\label{eq:Pinrelnonrel}
\begin{equation}
  \Pi^{(n)\text{r}}(q,\omega) = \left(1+\frac{\omega}{m_N}\right)^{n+1}
    \Pi^{(n)\text{nr}}(q,\omega(1+\omega/2m_N)) .
\label{eq:Pinrelomega}
\end{equation}
Actually, all the response functions derived below are expressed in terms of a 
generic scaling variable $\psi$, as $\Pi^{(n)}(q,\psi)$: One can then get the 
non-relativistic response
by using the (exact) expression (\ref{eq:psiNR}) for $\psi$ and the relativistic
response by using the (approximate) form (\ref{eq:psiR}) and multiplying each
polarization propagator by the appropriate power of $1+\omega/m_N$, i.~e.
\begin{equation}
  \Pi^{(n)\text{r}}(q,\omega) = \left(1+\frac{\omega}{m_N}\right)^{n+1}
    \Pi^{(n)\text{nr}}(q,\psi_{\text{r}}) .
\label{eq:Pinrelpsi}
\end{equation}
\end{mathletters}
Note that in the calculations of Refs.~\cite{Bar96a,Bar96b} only an overall
Jacobian factor, $1+\omega/m_N$, has been applied to the RPA response functions.

\subsection{ Free response }
\label{subsec:free-resp}

Although the free Fermi gas response function is a subject for textbooks (see,
e.~g., Ref.~\cite{Fet71}), it is useful to derive it here using a slightly
different approach, since it illustrates at the simplest level the method we
have adopted to overcome a major technical difficulty one meets in nuclear
matter calculations, --- namely the presence of $\theta$ functions, which
considerably complicate analytic integrations. 
As a side effect, also the calculation of $\Pi^{(0)}$ comes out much more
compact than in standard treatments.

From Eqs.~(\ref{eq:PiLT}) and (\ref{eq:OLT}), one immediately finds that 
\begin{equation}
  \Pi^{(0)}_{\text{L};I=0} = \Pi^{(0)}_{\text{L};I=1} =
  \Pi^{(0)}_{\text{T};I=0} = \Pi^{(0)}_{\text{T};I=1} \equiv
  \Pi^{(0)} ,
\end{equation}
where, following (\ref{eq:Gph}) and (\ref{eq:Pi}) we have defined 
\begin{equation}
  \Pi^{(0)}(q,\omega) = \int\frac{d\bbox{k}}{(2\pi)^3} 
    G^{(0)}_{\text{ph}}(\bbox{k},\bbox{q};\omega) ,
\label{eq:Pi0G0}
\end{equation}
having set
\begin{equation}
  G^{(0)}_{\text{ph}}(\bbox{k},\bbox{q};\omega) = 
    -i \int\frac{dk_0}{2\pi}G^{(0)}(\bbox{k}+\bbox{q},k_0+\omega)
    G^{(0)}(k,k_0) ,
\label{eq:G0ph}
\end{equation}
$G^{(0)}(k,k_0)$ being the free one-body propagator
\begin{equation}
  G^{(0)}(k,k_0) = 
    \frac{\theta(k-k_F)}{k_0-\epsilon^{(0)}_{\bbox{k}}+i\eta} +
    \frac{\theta(k_F-k)}{k_0-\epsilon^{(0)}_{\bbox{k}}-i\eta} .
\end{equation}
The integration over $k_0$ in (\ref{eq:G0ph}) is straightforward, yielding
\begin{equation}
  G^{(0)}_{\text{ph}}(\bbox{k},\bbox{q};\omega) = 
    \frac{\theta(k_F-k)\theta(|\bbox{k}+\bbox{q}|-k_F)}
    {\omega-\epsilon^{(0)}_{\bbox{k}+\bbox{q}}+\epsilon^{(0)}_{\bbox{k}}+i\eta}+
    \frac{\theta(k-k_F)\theta(k_F-|\bbox{k}+\bbox{q}|)}
    {-\omega+\epsilon^{(0)}_{\bbox{k}+\bbox{q}}-\epsilon^{(0)}_{\bbox{k}}+i\eta}
    ,
\label{eq:G0phold}
\end{equation}
which, inserted back into (\ref{eq:Pi0G0}), would give the standard definition 
of $\Pi^{(0)}$. Instead, let us rewrite $G^{(0)}_{\text{ph}}$ as
\begin{eqnarray}
  G^{(0)}_{\text{ph}}(\bbox{k},\bbox{q};\omega) &=&
    \frac{\theta(k_F-k)\theta(|\bbox{k}+\bbox{q}|-k_F)}
    {\omega-\epsilon^{(0)}_{\bbox{k}+\bbox{q}}+\epsilon^{(0)}_{\bbox{k}}+i\eta}+
    \frac{\theta(k-k_F)\theta(k_F-|\bbox{k}+\bbox{q}|)}
    {-\omega+\epsilon^{(0)}_{\bbox{k}+\bbox{q}}-\epsilon^{(0)}_{\bbox{k}}+i\eta}
    \nonumber \\
  && + \frac{\theta(k_F-k)\theta(k_F-|\bbox{k}+\bbox{q}|)}
    {\omega-\epsilon^{(0)}_{\bbox{k}+\bbox{q}}+\epsilon^{(0)}_{\bbox{k}}
      +i\eta_{\omega}} +
    \frac{\theta(k_F-k)\theta(k_F-|\bbox{k}+\bbox{q}|)}
    {-\omega+\epsilon^{(0)}_{\bbox{k}+\bbox{q}}-\epsilon^{(0)}_{\bbox{k}}
      -i\eta_{\omega}} ,
\end{eqnarray}
having added and subtracted the quantity in the second line, where we have set 
$\eta_{\omega}=\text{sign}(\omega)\eta$. A few algebraic manipulations then
yield
\begin{equation}
  G^{(0)}_{\text{ph}}(\bbox{k},\bbox{q};\omega) =
    \frac{\theta(k_F-k)-\theta(k_F-|\bbox{k}+\bbox{q}|)}
    {\omega-\epsilon^{(0)}_{\bbox{k}+\bbox{q}}+\epsilon^{(0)}_{\bbox{k}}
      +i\eta_{\omega}} .
\label{eq:G0phnew}
\end{equation}
Hence, from (\ref{eq:Pi0G0}) one gets
\begin{eqnarray}
  \Pi^{(0)}(q,\omega) &=& \int\frac{d\bbox{k}}{(2\pi)^3}\theta(k_F-k)
    \left[\frac{1}
    {\omega-\epsilon^{(0)}_{\bbox{k}+\bbox{q}}+\epsilon^{(0)}_{\bbox{k}}
      +i\eta_{\omega}} +
    \frac{1}
      {-\omega-\epsilon^{(0)}_{\bbox{k}+\bbox{q}}+\epsilon^{(0)}_{\bbox{k}}
      -i\eta_{\omega}}\right] \nonumber \\
  &=& \frac{m_N}{q}\frac{k_F^2}{(2\pi)^2}
    \left[{\cal Q}^{(0)}(\psi)-{\cal Q}^{(0)}(\psi+\bar{q})\right] .
\label{eq:Pi0}
\end{eqnarray}
Note that only one $\theta$ function forcing $k$ below $k_F$ is left, Pauli
blocking being enforced by cancellations between the energy denominators.
In (\ref{eq:Pi0}), we have introduced $\bar{q}=q/k_F$ and the adimensional 
function
\begin{equation}
  {\cal Q}^{(0)}(\psi) =
  \frac{1}{2}\int_{-1}^{1}dy\frac{1-y^2}{\psi-y+i\eta_{\omega}} ,
\end{equation}
which is easily evaluated, yielding 
\begin{mathletters}
\begin{eqnarray}
  \text{Re}{\cal Q}^{(0)}(\psi) &=& \psi+\frac{1}{2}(1-\psi^2)
    \text{ln}\left|\frac{1+\psi}{1-\psi}\right| = 
    \frac{2}{3}[Q_0(\psi)-Q_2(\psi)] \\
  \text{Im}{\cal Q}^{(0)}(\psi) &=& -\text{sign}(\omega)\theta(1-\psi^2)
    \frac{\pi}{2}(1-\psi^2) = -\text{sign}(\omega)\theta(1-\psi^2)
    \frac{\pi}{3}[P_0(\psi)-P_2(\psi)] ,
\end{eqnarray}
\end{mathletters}
where $P_n$ and $Q_n$ are Legendre polynomials and Legendre functions of second
kind, respectively.

\subsection{ Hartree-Fock response }
\label{subsec:HF-resp}

The HF polarization propagator in nuclear matter is obtained by dressing the
one-body propagators appearing in $\Pi^{(0)}$ with the first order self-energy
$\Sigma^{(1)}$, so that one can follow essentially the same derivation of the 
previous subsection. The spin-isospin matrix elements are the same as for the 
free response, yielding
\begin{equation}
  \Pi^{\text{HF}}_{\text{L};I=0} = \Pi^{\text{HF}}_{\text{L};I=1} =
  \Pi^{\text{HF}}_{\text{T};I=0} = \Pi^{\text{HF}}_{\text{T};I=1} \equiv
  \Pi^{\text{HF}} ,
\end{equation}
where 
\begin{equation}
  \Pi^{\text{HF}}(q,\omega) = \int\frac{d\bbox{k}}{(2\pi)^3}\theta(k_F-k)
    \left[\frac{1}
    {\omega-\epsilon^{(1)}_{\bbox{k}+\bbox{q}}+\epsilon^{(1)}_{\bbox{k}}
      +i\eta_{\omega}} +
    \frac{1}
    {-\omega-\epsilon^{(0)}_{\bbox{k}+\bbox{q}}+\epsilon^{(0)}_{\bbox{k}}
      -i\eta_{\omega}}\right] ,
\label{eq:PiHF}
\end{equation}
with $ \epsilon^{(1)}_{\bbox{k}} = \epsilon^{(0)}_{\bbox{k}} +
    \Sigma^{(1)}(k) $.

Although the evaluation of the HF response is numerically quite straightforward,
in Ref.~\cite{Bar96a} an analytic approximation for $\text{Im}\Pi^{\text{HF}}$
has been worked out, with the aim of using it to include HF correlations in RPA
calculations. Here, it will be shown that the validity of that approximation is
more general, not being limited to the HF response, although in the latter case
one can directly check the good accuracy of the procedure. 

In any Feynman diagram considered here and in the following, the nucleon
self-energy enters through the ph energy denominators,
\begin{equation}
  \omega-\epsilon^{(1)\text{nr}}_{\bbox{k}+\bbox{q}}+
    \epsilon^{(1)\text{nr}}_{\bbox{k}} =
    \omega - \frac{(\bbox{k}+\bbox{q})^2}{2m_N} + \frac{k^2}{2m_N} 
    -\Sigma^{(1)}(|\bbox{k}+\bbox{q}|)+\Sigma^{(1)}(k) ,
\label{eq:endenSE}
\end{equation}
where the non-relativistic expression for the nucleon kinetic energy has been
used. In (\ref{eq:endenSE}), one can assume to have always $k<k_F$ and 
$|\bbox{k}+\bbox{q}|>k_F$. Although the latter may not look immediately apparent
from, e.~g., Eq.~(\ref{eq:PiHF}), remember that cancellations between the energy
denominators are such to enforce the Pauli principle; the same
will also be true for the RPA diagrams\footnote{ It should also be noted that
the infinite Fermi gas is better suited for relatively large momenta 
($q\gtrsim 2k_F$), where the conditions above are satisfied by definition.}.

Clearly, if $\Sigma^{(1)}(k)$ were parabolic in the momentum, the inclusion of
the self-energy would be achieved simply by substituting $m_N$ with an effective
mass. For realistic potentials, a parabolic fit of the self-energy over the
whole range of momenta is not, in general, a good approximation. 
It is a good approximation, on the other hand, to fit {\em separately} the
particle and the hole part of the self-energy, restricting the fit to the range
of momenta actually involved in the integration.
Since in Eq.~(\ref{eq:PiHF}) (but also in the RPA diagrams discussed later) one
has $k$ integrated from 0 to $k_F$ and, furthermore, $|\bbox{k}+\bbox{q}|>k_F$,
one can set
\begin{eqnarray}
  \Sigma^{(1)} &\cong& \bar{A} + \bar{B}\frac{k^2}{2m_N} ,
    \quad 0<k<k_F , \nonumber \\
    \\
  \Sigma^{(1)} &\cong& A + B\frac{k^2}{2m_N} ,
    \quad \text{max}(q-k_F,k_F)<k<q+k_F . \nonumber
\end{eqnarray}
Inserting this ``biparabolic approximation'' back into (\ref{eq:endenSE}) 
and setting $\varepsilon=\bar{A}-A$ one gets 
\begin{equation}
  \omega-\epsilon^{(1)\text{nr}}_{\bbox{k}+\bbox{q}}+
    \epsilon^{(1)\text{nr}}_{\bbox{k}} \cong
  \frac{qk_F}{m_N^{*\text{nr}}}\left[
    \psi_{\text{nr}}^*-\hat{\bbox{q}}\cdot\frac{\bbox{k}}{k_F} \right] ,
\label{eq:endenSEappr}
\end{equation}
where we have
neglected a term proportional to $k^2$. It can be expected to be small, since
$k<k_F$ and, typically, $q>k_F$; it has, however, to be checked for any given
interaction, since it involves the biparabolic fit parameters $B$ and $\bar{B}$.
In Ref.~\cite{Bar96a} it had been shown to be small for the Bonn potential; the
same turns out to be true also for the effective interaction employed in the
next section.

Eq.~(\ref{eq:endenSEappr}) is similar to the expression (\ref{eq:endenNR}) for
the free energy denominator, but for the substitutions
\begin{eqnarray}
  m_N &\to& m_N^{*\text{nr}} = \frac{m_N}{1+B} \nonumber \\
  \psi_{\text{nr}} &\to& \psi_{\text{nr}}^* = 
    \frac{1}{k_F}\left[
    (\omega+\varepsilon)\frac{m_N^*}{q}-\frac{q}{2}\right] 
    \label{eq:mNstar} \\
  && \phantom{\psi_{\text{nr}}^*} = \frac{\psi_{\text{nr}}+\chi}{1+B} ,
    \qquad \chi = \frac{1}{k_F}
      \left(\frac{\varepsilon m_N}{q}-B\frac{q}{2}\right) ,
    \nonumber
\end{eqnarray}
(or $\omega\to\omega+\varepsilon$).

In Ref.~\cite{Bar96a} relativistic kinematics had been accounted for by applying
the transformation $\omega\to\omega(1+\omega/2m_N)$ discussed above to the
previous formulae. The correct approximation can be worked out by starting again
from the ph propagator and rewriting it as 
($\Delta\Sigma^{(1)}(\bbox{k},\bbox{q})\equiv
\Sigma^{(1)}(k)-\Sigma^{(1)}(|\bbox{k}+\bbox{q}|)$)
\begin{eqnarray}
  && \frac{1}{\omega-\epsilon^{(1)\text{r}}_{\bbox{k}+\bbox{q}}+
    \epsilon^{(1)\text{r}}_{\bbox{k}}} = \nonumber \\
  && \quad =
    \frac{\omega+\sqrt{k^2+m_N^2}+\Delta\Sigma^{(1)}(\bbox{k},\bbox{q})+
    \sqrt{(\bbox{k}+\bbox{q})^2+m_N^2}}{\omega^2+2\omega\sqrt{k^2+m_N^2}+
    2(\omega+\sqrt{k^2+m_N^2})\Delta\Sigma^{(1)}(\bbox{k},\bbox{q})+
    [\Delta\Sigma^{(1)}(\bbox{k},\bbox{q})]^2-q^2-2\bbox{q}\cdot\bbox{k}} 
    \nonumber \\
  && \quad \cong 
    \frac{m_N^{*\text{r}}}{q k_F}\frac{1+\omega/m_N+\Delta^{(1)}/m_N}
    {\psi^*_{\text{r}}-\hat{\bbox{q}}\cdot\bbox{k}/k_F} ,
\label{eq:endenSErel}
\end{eqnarray}
where
\begin{eqnarray}
  m_N^{*\text{r}} &=& \frac{m_N}{1+B(1+\omega/m_N+\Delta^{(1)}/m_N)} 
    \nonumber \\
  \psi^*_{\text{r}} &=& \frac{\psi_{\text{r}}+
  \chi[1+B(1+\omega/m_N+\Delta^{(1)}/2m_N)]}{1+B(1+\omega/m_N+\Delta^{(1)}/m_N)}
  \label{eq:mNstarrel} \\
  \Delta^{(1)} &=& \varepsilon-B\frac{q^2}{2m_N}\equiv\frac{q k_F}{m_N}\chi , 
    \nonumber
\end{eqnarray}
with $\chi$ already defined in (\ref{eq:mNstar}).
In deriving (\ref{eq:endenSErel}), we have assumed $k^2<<m_N^2$, evaluated the 
numerator at the pole discarding then any angular dependence and, in the 
denominator, retained only terms at most linear in 
$\hat{\bbox{q}}\cdot\bbox{k}/k_F$.
As one can see, beside the transformation $\omega\to\omega(1+\omega/2m_N)$ there
are other relativistic corrections, both to the effective scaling variable and 
to the Jacobian.

The quality of the approximations introduced above is good, with at most a few 
per cent discrepancy (except on the borders of the response region, where the 
Fermi gas is anyway unrealistic).
Thus, we see that either in the non-relativistic or relativistic case, the 
prescription to include HF correlations in a response function is 
simply to replace $\psi$ with $\psi^*$ and $m_N$ with $m_N^*$ (and to multiply
by a normalization factor when employing relativistic kinematics (see 
Eqs.~(\ref{eq:Pinrelnonrel})).
For instance, from (\ref{eq:Pi0}) one gets
\begin{equation}
  \Pi^{\text{HF}}(q,\omega) \cong J \frac{m_N^*}{q}\frac{k_F^2}{(2\pi)^2}
    \left[{\cal Q}^{(0)}(\psi^*)-{\cal Q}^{(0)}(\psi^*+\bar{q})\right] ,
\label{eq:HFappr}
\end{equation}
with
\begin{eqnarray}
  J_{\text{nr}} &=& 1 \nonumber \\
  \label{eq:JHF} \\
  J_{\text{r}}  &=& 1+\frac{\omega}{m_N}+\frac{\Delta^{(1)}}{m_N} . \nonumber
\end{eqnarray}

\subsection{ Random phase approximation response }
\label{subsec:RPA-resp}

\begin{figure}[t]
\begin{center}
\vskip 2mm
\mbox{\epsfig{file=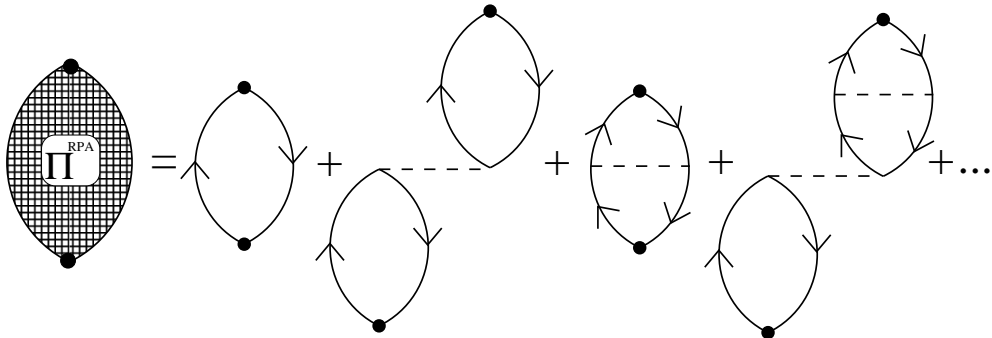,width=.8\textwidth}}
\caption{ Diagrammatic representation of the perturbative expansion for the
polarization propagator in random phase approximation.
 }
\label{fig:PiRPA}
\end{center}
\end{figure}

If in Eq.~(\ref{eq:Gph}) one substitutes the irreducible vertex function
$\Gamma^{13}$ with the matrix elements of the bare potential, one gets the so
called {\em random phase approximation} to $G^{\text{ph}}$. In terms of the
polarization propagator (\ref{eq:Pi}) one would get an infinite sum of diagrams
such as those of Fig.~\ref{fig:PiRPA}.

We have already noted at the beginning of Section~\ref{sec:Resp} that, while for
the two-body Green's function $G^{\text{ph}}$ one can introduce an
integral equation, this is not possible, in general, for the polarization
propagator. It becomes possible when one approximates the irreducible vertex 
function $\Gamma^{13}$ with the {\em direct} matrix elements of the interaction.
In that case, in an infinite system, one gets a simple algebraic equation, whose
solution, for the polarization propagators (\ref{eq:PiLT}) and the interaction
(\ref{eq:pot}), is readily found to be
\begin{equation}
  \Pi^{\text{ring}}_{\text{X}}(q,\omega) = \frac{\Pi^{(0)}(q,\omega)}
    {1-\Pi^{(1)\text{d}}_{\text{X}}(q,\omega)\big/\Pi^{(0)}(q,\omega)}
    ,
\label{eq:Piring}
\end{equation}
where $\Pi^{(1)\text{d}}_{\text{X}}$ represents the first order {\em direct}
polarization propagator:
\begin{mathletters}
\label{eq:Pi1d}
\begin{eqnarray}
  \Pi^{(1)\text{d}}_{\text{L};I=0(1)}(q,\omega) &=& 
    \Pi^{(0)}(q,\omega)4V_{0(\tau)}(q)\Pi^{(0)}(q,\omega) , \\
  \Pi^{(1)\text{d}}_{\text{T};I=0(1)}(q,\omega) &=&
    \Pi^{(0)}(q,\omega)4[V_{\sigma(\sigma\tau)}(q)
      -V_{t(t\tau)}(q)]\Pi^{(0)}(q,\omega) .
\end{eqnarray}
\end{mathletters}
The effect of the exchange diagrams is often included through an effective 
zero-range interaction, calculated by taking the limit $q\to0$ of the first
order exchange contribution and rewriting it as an effective first order direct
term \cite{Ose82}. Exact calculations, however, show that extrapolating this
approximation to finite transferred momenta is not always reliable \cite{Shi89}.

A more sophisticated approximation scheme is given by the {\em continued
fraction} {\em expansion} \cite{Len80,Del85,Del87,Fes92}.
At infinite order the CF expansion gives the exact result as the summation of
the perturbative series, so that it is not easier to calculate: However, when
truncated at finite order, it reproduces the standard perturbative series at the
same order plus an approximation for each one of the infinite number of higher 
order contributions. The trouble here is that there is no general method to 
predict the convergence of the CF expansion, the only reliable test being a 
direct comparison of the results at successive orders.

On the other hand, we should note that for zero-range forces the first order CF
expansion already gives the exact (albeit trivial) result, making one hope that
the short-range nature of the nuclear interactions allows for a fast
convergence. Indeed, all available calculations have been performed truncating
the CF expansion at first order \cite{Del85,Del87,Alb93,Bar94,Bar96a,Bar96b}.
Here, as anticipated, we shall test the convergence up to second order.

The CF formalism for the polarization propagator is developed in 
Ref.~\cite{Del85} for the case of Tamm-Dancoff correlations and extended in 
Ref.~\cite{Del87} to the full RPA. Instead of following the rather involved
formal derivation given there, we shall briefly sketch a sort of heuristic
derivation of the CF expansion (which is, of course, only possible ``a
posteriori'', once the meaning of the CF series has been understood).

Let us assume that we want to build a CF-like expansion for the polarization
propagator, according to the pattern 
\begin{equation}
  \Pi^{\text{RPA}} = \frac{\Pi^{(0)}}
    {\displaystyle 1 - A - \frac{\strut B}{\displaystyle 1 - C - \frac{\strut D}
    {\displaystyle 1 - ...}}} .
\end{equation}
We have said that the CF approach at $n$-th order reproduces the perturbative 
series at the same order and then it approximates higher orders.
Thus, if we want to approximate at first order in CF the exact RPA propagator
(for sake of illustration we drop spin-isospin indices),
\begin{equation}
  \Pi^{\text{RPA}} = \sum_{n=0}^{\infty}\Pi^{(n)} ,
\end{equation}
we can rather naturally write 
\begin{equation}
  \Pi^{(n)} \cong \Pi^{(0)} \left[\frac{\Pi^{(1)}}{\Pi^{(0)}}\right]^n ,
\end{equation}
where $\Pi^{(1)}\equiv\Pi^{(0)}4V\Pi^{(0)}+\Pi^{(1)\text{ex}}$ is the sum of the
direct and exchange first order terms, --- since this is the correct expression
for the direct terms. 
With this approximation the summation is trivial, yielding
\begin{equation}
  \Pi^{\text{RPA}}_{\text{CF1}} = 
    \frac{\Pi^{(0)}}{1-\Pi^{(1)}/\Pi^{(0)}} =
    \frac{\Pi^{(0)}}{1-4V\Pi^{(0)}-\Pi^{(1)\text{ex}}/\Pi^{(0)}} .
\end{equation}
We could then add in the denominator of the expression above the exact second
order term, $\Pi^{(2)}$, having care of subtracting the approximation to it
provided by the first order CF expansion, $[\Pi^{(1)}]^2/\Pi^{(0)}$.
Then, we would get
\begin{eqnarray}
  \Pi^{\text{RPA}}_{\text{CF2}} &=& 
    \frac{\Pi^{(0)}}{1-\Pi^{(1)}/\Pi^{(0)}
    -\{\Pi^{(2)}/\Pi^{(0)}-[\Pi^{(1)}/\Pi^{(0)}]^2\}} \nonumber \\
  &=& \frac{\Pi^{(0)}}{1-4V\Pi^{(0)}-\Pi^{(1)\text{ex}}/\Pi^{(0)}
    -\{\Pi^{(2)\text{ex}}/\Pi^{(0)}-[\Pi^{(1)\text{ex}}/\Pi^{(0)}]^2\}} .
\label{eq:PiCF2}
\end{eqnarray}
From Eq.~(\ref{eq:PiCF2}) it is easy to check that the third order term is
approximated as $\Pi^{(3)}\cong\Pi^{(1)}\{2\Pi^{(2)}/\Pi^{(0)}-
[\Pi^{(1)}/\Pi^{(0)}]^2\}$.
Then, going ahead in a CF-style expansion we would guess for the exact RPA
propagator the following expression:
\begin{equation}
  \Pi^{\text{RPA}} = \frac{\Pi^{(0)}}
    {\displaystyle 1 - \Pi^{(1)}/\Pi^{(0)}
    - \frac{\strut 
      \Pi^{(2)\text{ex}}/\Pi^{(0)}-[\Pi^{(1)\text{ex}}/\Pi^{(0)}]^2}
    {\displaystyle 1 - 
      \frac{\strut \Pi^{(3)\text{ex}}/\Pi^{(0)}+[\Pi^{(1)\text{ex}}/\Pi^{(0)}]^3
        -2[\Pi^{(1)\text{ex}}/\Pi^{(0)}][\Pi^{(2)\text{ex}}/\Pi^{(0)}]}
      {\displaystyle 
      \Pi^{(2)\text{ex}}/\Pi^{(0)}-[\Pi^{(1)\text{ex}}/\Pi^{(0)}]^2}
    - ...}} .
\end{equation}
This is exactly the expression one would get from the formalism of 
Refs.~\cite{Del85,Del87} if one had the patience to work out the expansion up to
third order. Note that we did not assume any specific scheme (either
Tamm-Dancoff or RPA) in this heuristic derivation.

Thus, following (\ref{eq:Piring}) we can write
\begin{equation}
  \Pi^{\text{RPA}}_{\text{X}} = \frac{\Pi^{(0)}}
    {\displaystyle 1 
    -\Pi^{(1)\text{d}}_{\text{X}}\big/\Pi^{(0)}
    -\Pi^{(1)\text{ex}}_{\text{X}}\big/\Pi^{(0)}
    -\frac{\strut 
      \Pi^{(2)\text{ex}}_{\text{X}}\big/\Pi^{(0)}
      -\left[\Pi^{(1)\text{ex}}_{\text{X}}\big/\Pi^{(0)}
      \right]^2}{\displaystyle 1-...}} ,
\end{equation}
where $\Pi^{(1)\text{d}}_{\text{X}}$ has been defined in (\ref{eq:Pi1d}).
Then, truncation at $n$-th order requires the calculation of the exchange
contributions up to that order.

From (\ref{eq:PiLT}) we can write
\begin{mathletters}
\begin{eqnarray}
  \Pi^{(n)\text{ex}}_{\text{L};I}(q,\omega) &=& 
    \text{tr}[\hat{O}^{I}_{\text{L}}\hat{\Pi}^{(n)\text{ex}}(q,\omega)
    \hat{O}^{I}_{\text{L}}] = 
    \sum_{\alpha_i}
    C^{\alpha_1...\alpha_n}_{\text{L};I}\Pi^{(n)\text{ex}}_{\alpha_1...\alpha_n}
    (q,\omega)
    , \\
  \Pi^{(n)\text{ex}}_{\text{T};I}(q,\omega) &=&
    \sum_{ij}\Lambda_{ji}
    \text{tr}[\hat{O}^{I}_{\text{T};i}\hat{\Pi}^{(n)\text{ex}}(q,\omega)
    \hat{O}^{I}_{\text{T};j}] = 
    \sum_{\alpha_i}
    C^{\alpha_1...\alpha_n}_{\text{T};I}\Pi^{(n)\text{ex}}_{\alpha_1...\alpha_n}
    (q,\omega)
    , 
\end{eqnarray}
\end{mathletters}
where the indices $\alpha_i$ run over all the spin-isospin channels and the 
spin-isospin factors are condensed in the coefficients 
$C^{\alpha_1...\alpha_n}_{\text{X}} \equiv C^{(\alpha_1)}_{\text{X}} 
C^{(\alpha_2)}_{\text{X}}...C^{(\alpha_n)}_{\text{X}}$ (see Table~\ref{tab:I}).
\begin{table}
\begin{center}
\begin{tabular}{crrrrrr}
  $\strut\phantom{\Big|}X$ & $C^0_X$ & $C^\tau_X$ & $C^\sigma_X$ & 
  $C^{\sigma\tau}_X$ & $C^t_X$ & $C^{t\tau}_X$ \\ \tableline
  $L;I=0$ & 1  &  3  &  3  &  9  &  0  &  0  \\ 
  $L;I=1$ & 1  & -1  &  3  & -3  &  0  &  0  \\ 
  $T;I=0$ & 1  &  3  & -1  & -3  & -1  & -3  \\ 
  $T;I=1$ & 1  & -1  & -1  &  1  & -1  &  1  \\ 
\end{tabular}
\end{center}
\caption{ The spin-isospin coefficients $C^{\alpha}_X$ (see text), in the 
longitudinal and transverse isoscalar and isovector channels, for the 
interaction (\protect\ref{eq:pot}).
  }
\label{tab:I}
\end{table}
We have introduced the ``elementary'' exchange contribution
$\Pi^{(n)\text{ex}}_{\alpha_1...\alpha_n}$ containing $n$ interaction lines
$V_{\alpha_1}$...$V_{\alpha_n}$, namely\footnote{ The following formulae are
valid for non-tensor interactions; the treatment of the tensor terms is slightly
more complex and it is given in Appendix~\ref{app:C}.}
\begin{eqnarray}
  \Pi^{(n)\text{ex}}_{\alpha_1...\alpha_n}(q,\omega) &=&
    -i^{n+1}\int\frac{d^4K_1}{(2\pi)^4}\cdot\cdot\cdot
    \frac{d^4K_{n+1}}{(2\pi)^4}
    G^{(0)}(K_1)G^{(0)}(K_1+Q)V_{\alpha_1}(\bbox{k}_1-\bbox{k}_2)
    \cdot\cdot\cdot \nonumber \\
  &&\qquad \cdot\cdot\cdot V_{\alpha_n}(\bbox{k}_n-\bbox{k}_{n+1})
    G^{(0)}(K_{n+1})G^{(0)}(K_{n+1}+Q) \nonumber \\
  &=& (-1)^n\int\frac{d\bbox{k}_1}{(2\pi)^3}\cdot\cdot\cdot
    \frac{d\bbox{k}_{n+1}}{(2\pi)^3}
    G^{(0)}_{\text{ph}}(\bbox{k}_1,\bbox{q};\omega)
    V_{\alpha_1}(\bbox{k}_1-\bbox{k}_2)\cdot\cdot\cdot \nonumber \\
  &&\qquad \cdot\cdot\cdot V_{\alpha_n}(\bbox{k}_n-\bbox{k}_{n+1})
    G^{(0)}_{\text{ph}}(\bbox{k}_{n+1},\bbox{q};\omega) .
\end{eqnarray}
With the definition of $G^{(0)}_{\text{ph}}$ given in (\ref{eq:G0phnew})
one can, --- by a suitable change of integration variables, --- eliminate all
the $\theta$-functions that contain angular integration variables, leaving a 
multiple integral with the following general structure:
\begin{eqnarray}
  && \Pi^{(n)\text{ex}}_{\alpha_1...\alpha_n}(q,\omega) =
    (-1)^n\int\frac{d\bbox{k}_1}{(2\pi)^3}\theta(k_F-k_1)
    \cdot\cdot\cdot
    \frac{d\bbox{k}_{n+1}}{(2\pi)^3}\theta(k_F-k_{n+1}) \nonumber \\
  &&\quad\times\Big[
    \frac{1}{\omega-\epsilon_{\bbox{k}_1+\bbox{q}}+\epsilon_{\bbox{k}_1}
    +i\eta_\omega} 
    V_{\alpha_1}(\bbox{k}_1-\bbox{k}_2)\cdot\cdot\cdot
    V_{\alpha_n}(\bbox{k}_n-\bbox{k}_{n+1})
    \frac{1}{\omega-\epsilon_{\bbox{k}_{n+1}+\bbox{q}}+\epsilon_{\bbox{k}_{n+1}}
    +i\eta_\omega} \nonumber \\
  &&\quad\quad+\sum(\omega\to-\omega) \Big] .
\label{eq:Pinex}
\end{eqnarray}
In (\ref{eq:Pinex}), $\sum(\omega\to-\omega)$ stands for the sum of all the 
terms generated according to the following rules:
\begin{itemize}
  \item[i)] take all the terms generated by substituting, in the second line 
of (\ref{eq:Pinex}), in one energy denominator $\omega\to-\omega$; then, by 
doing the same substitution in two energy denominators; and so on up to making 
the replacement $\omega\to-\omega$ in all the $n+1$ denominators;
  \item[ii)] every time 
$(\omega-\epsilon_{\bbox{k}_i+\bbox{q}}+\epsilon_{\bbox{k}_i}
  +i\eta_\omega)^{-1}$ is replaced with
$(-\omega-\epsilon_{\bbox{k}_i+\bbox{q}}+\epsilon_{\bbox{k}_i}
  -i\eta_\omega)^{-1}$,
then replace $\bbox{k}_i$ with $-\bbox{k}_i-\bbox{q}$ in the potentials.
\end{itemize}

The number of integrations can be reduced by noticing that the azimuthal angles
are contained only in the potential functions $V_{\alpha_i}$.
For typical nuclear physics potentials this integration can be done 
analytically, so that it is convenient to introduce a new function representing
the azimuthal integral of the potential. To this end, let us define new 
variables:
\begin{eqnarray}
  |\bbox{k}-\bbox{k}'| &=& \sqrt{k^2+{k'}^2-2kk'[\cos\theta\,\cos\theta'+
    \sin\theta\,\sin\theta'\,\cos(\varphi-\varphi')]} \nonumber \\
  &=& \sqrt{k^2+{k'}^2-2[yy'+\sqrt{k^2-y^2}\sqrt{{k'}^2-{y'}^2}
    \cos(\varphi-\varphi')]} \\
  &=& \sqrt{x+x'-2\sqrt{x}\sqrt{x'}\cos(\varphi-\varphi')+(y-y')^2} , 
    \nonumber
\end{eqnarray}
where $y\equiv k\,\cos\theta$ and $x\equiv k^2-y^2$.
Then, we can introduce
\begin{equation}
  W_{\alpha}(x,y;x',y') = \int_0^{2\pi}\frac{d\varphi}{2\pi}
    V_{\alpha}(\bbox{k}-\bbox{k}') = W_{\alpha}(x',y';x,y)
\label{eq:Walpha}
\end{equation}
and rewrite (\ref{eq:Pinex}) as 
\begin{eqnarray}
  \Pi^{(n)\text{ex}}_{\alpha_1...\alpha_n}(q,\omega) &=&
    (-1)^n\left(\frac{m_N}{q}\right)^{n+1}\left(\frac{k_F}{2\pi}\right)^{2n+2}
    \int_{-1}^{1}dy_1\frac{1}{2}\int_{0}^{1-y_1^2}dx_1\cdot\cdot\cdot
    \int_{-1}^{1}dy_{n+1}\frac{1}{2}\int_{0}^{1-y_{n+1}^2}dx_{n+1} \nonumber \\
  && \times\frac{1}{\psi-y_1+i\eta_{\omega}}W_{\alpha_1}(x_1,y_1;x_2,y_2)
   \cdot\cdot\cdot W_{\alpha_n}(x_n,y_n;x_{n+1},y_{n+1})
   \frac{1}{\psi-y_{n+1}+i\eta_{\omega}} \nonumber \\
  && + \sum(\omega\to-\omega) .
\label{eq:Pinexpsi}
\end{eqnarray}
For $n=1$ one has
\begin{eqnarray}
  \Pi^{(1)\text{ex}}_{\alpha}(q,\omega) &=&
    -\left(\frac{m_N}{q}\right)^{2}\frac{k_F^4}{(2\pi)^4}
    \int_{-1}^{1}dy\frac{1}{2}\int_{0}^{1-y^2}dx
    \int_{-1}^{1}dy'\frac{1}{2}\int_{0}^{1-{y'}^2}dx' \nonumber \\
  && \times\frac{1}{\psi-y+i\eta_{\omega}}W_{\alpha}(x,y;x',y')
   \frac{1}{\psi-y'+i\eta_{\omega}} \nonumber \\
  && + \sum(\omega\to-\omega) \nonumber \\
  &=&  -\left(\frac{m_N}{q}\right)^{2}\frac{k_F^4}{(2\pi)^4}
    \left[{\cal Q}_\alpha^{(1)}(0,\psi) 
    - {\cal Q}_\alpha^{(1)}(\bar{q},\psi) 
    + {\cal Q}_\alpha^{(1)}(0,\psi+\bar{q})
    - {\cal Q}_\alpha^{(1)}(-\bar{q},\psi+\bar{q})\right]
    , \nonumber \\
\label{eq:Pi1ex}
\end{eqnarray}
where 
\begin{equation}
  {\cal Q}_\alpha^{(1)}(\bar{q},\psi) = 2 \int_{-1}^1 dy
    \frac{1}{\psi-y+i\eta_{\omega}}\int_{-1}^1 dy' \, {W_\alpha}''(y,y';\bar{q})
    \frac{1}{y-y'+\bar{q}}
\label{eq:Q1alpha}
\end{equation}
and
\begin{equation}
  W''_\alpha(y,y';\bar{q}) = \frac{1}{2}\int_0^{1-y^2}dx\,
    \frac{1}{2}\int_0^{1-{y'}^2}dx' \, W_\alpha(x,y+\bar{q};x',y') .
\label{eq:Wppalpha}
\end{equation}
Note that in getting to Eq.~(\ref{eq:Q1alpha}) use has been made of the 
Poincar\'e--Bertrand theorem \cite{Bal63}. 
For the potential (\ref{eq:mes-exch}) $W''_\alpha$ can be calculated 
analytically (see Appendix~\ref{app:D}), so that the calculation of the first 
order exchange contribution to the polarization propagator is reduced to the 
numerical evaluation of two-dimensional integrals, --- for the real part, --- 
and of one-dimensional integrals, --- for the imaginary part.

For $n=2$ one has
\begin{eqnarray}
  &&\Pi^{(2)\text{ex}}_{\alpha\alpha'}(q,\omega) =
    \left(\frac{m_N}{q}\right)^{3}\frac{k_F^6}{(2\pi)^6}
    \int_{-1}^{1}dy_1\frac{1}{2}\int_{0}^{1-y_1^2}dx_1
    \int_{-1}^{1}dy_2\frac{1}{2}\int_{0}^{1-y_2^2}dx_2
    \int_{-1}^{1}dy_3\frac{1}{2}\int_{0}^{1-y_3^2}dx_3 \nonumber \\
  && \phantom{\Pi^{(2)\text{ex}}_{\alpha\alpha'}(q,\omega)=}
    \times\frac{1}{\psi-y_1+i\eta_{\omega}}W_{\alpha}(x_1,y_1;x_2,y_2)
    \frac{1}{\psi-y_2+i\eta_{\omega}}W_{\alpha}(x_2,y_2;x_3,y_3)
    \frac{1}{\psi-y_3+i\eta_{\omega}} \nonumber \\
  && \phantom{\Pi^{(2)\text{ex}}_{\alpha\alpha'}(q,\omega)=}  
    + \sum(\omega\to-\omega) \nonumber \\
  && \phantom{\Pi^{(2)\text{ex}}_{\alpha\alpha'}(q,\omega)} =
    \left(\frac{m_N}{q}\right)^{3}\frac{k_F^6}{(2\pi)^6}
    \left[{\cal Q}_{\alpha\alpha'}^{(2)}(0,0;\psi) 
        - {\cal Q}_{\alpha\alpha'}^{(2)}(0,\bar{q};\psi) 
        - {\cal Q}_{\alpha\alpha'}^{(2)}(\bar{q},0;\psi) 
        + {\cal Q}_{\alpha\alpha'}^{(2)}(\bar{q},\bar{q};\psi) \right.
    \nonumber \\
  &&\qquad \left.
        - {\cal Q}_{\alpha\alpha'}^{(2)}(0,0;\psi+\bar{q})
        + {\cal Q}_{\alpha\alpha'}^{(2)}(0,-\bar{q};\psi+\bar{q})
        + {\cal Q}_{\alpha\alpha'}^{(2)}(-\bar{q},0;\psi+\bar{q})
        - {\cal Q}_{\alpha\alpha'}^{(2)}(-\bar{q},-\bar{q};\psi+\bar{q})
    \right] , \nonumber \\
\label{eq:Pi2ex}
\end{eqnarray}
where 
\begin{equation}
  {\cal Q}_{\alpha\alpha'}^{(2)}(\bar{q}_1,\bar{q}_2;\psi) = \int_{-1}^1 dy
    \frac{1}{2}\int_{0}^{1-y^2} dx \, 
    {\cal G}_{\alpha}(x,y+\bar{q}_1;\psi+\bar{q}_1)
    \frac{1}{\psi-y+i\eta_{\omega}}
    {\cal G}_{\alpha'}(x,y+\bar{q}_2;\psi+\bar{q}_2)
\label{eq:Q2aap}
\end{equation}
and
\begin{mathletters}
\label{eq:GWp}
\begin{eqnarray}
  {\cal G}_{\alpha}(x,y;\psi) &=& \int_{-1}^1 dy'
    \frac{1}{\psi-y'+i\eta_{\omega}}W'_\alpha(x,y;y') ,
\label{eq:Galpha} \\
  W'_\alpha(x,y;y') &=& \frac{1}{2}\int_0^{1-{y'}^2}dx' \, W_\alpha(x,y;x',y')
    .
\label{eq:Wpalpha}
\end{eqnarray}
\end{mathletters}
For the potential (\ref{eq:mes-exch}) $W'_\alpha$ can be calculated analytically
(see Appendix~\ref{app:D}) and one is left with the numerical integration of 
(\ref{eq:Q2aap}) and (\ref{eq:Galpha}), so that the calculation of the second 
order exchange contribution to the polarization propagator is effectively 
reduced to the numerical evaluation of at most three-dimensional integrals.

Going to higher orders implies a numerical two-dimensional integration for each
supplemental interaction line, since, for a potential of the form 
(\ref{eq:mes-exch}), only the azimuthal integration can be performed 
analytically for the interaction lines that are not close to the external 
vertices.

The Hartree-Fock dressing of the nucleon propagators can again be done as 
explained in Subsection~\ref{subsec:HF-resp}, with the replacements 
$\psi\to\psi^*$ and $m_N\to m_N^*$, where $\psi^*$ and $m_N^*$ have been defined
in (\ref{eq:mNstar}) and (\ref{eq:mNstarrel}), multiplying by the correct power
of the normalization factor $1+\omega/m_N+\Delta^{(1)}/m_N$ when the 
relativistic kinematics is employed (see Eqs.~(\ref{eq:Pinrelnonrel}) and 
(\ref{eq:JHF}).

\section{ Results }
\label{sec:res}

First of all we have to choose the Fermi momentum. Of course, one could
easily perform a local density calculation to achieve a better description of
finite nuclei: Here, for sake of illustration, we prefer to use the pure
Fermi gas. The choice of $k_F$ can be done in several ways: We shall choose an
average value according to the formula
\begin{equation}
  \bar{k}_F = \frac{1}{A}\int d\bbox{r} k_F(r) \rho(r) ,
\end{equation}
where $\rho(r)$ is the empirical Fermi density distribution normalized to the
number of nucleons and $k_F(r)=[(3\pi/2)\rho(r)]^{1/3}$.
For $^{12}$C one gets $\bar{k}_F\approx195$ MeV/c and this is the value used in
the calculations that follow.

\begin{figure}[p]
\begin{center}
\mbox{\epsfig{file=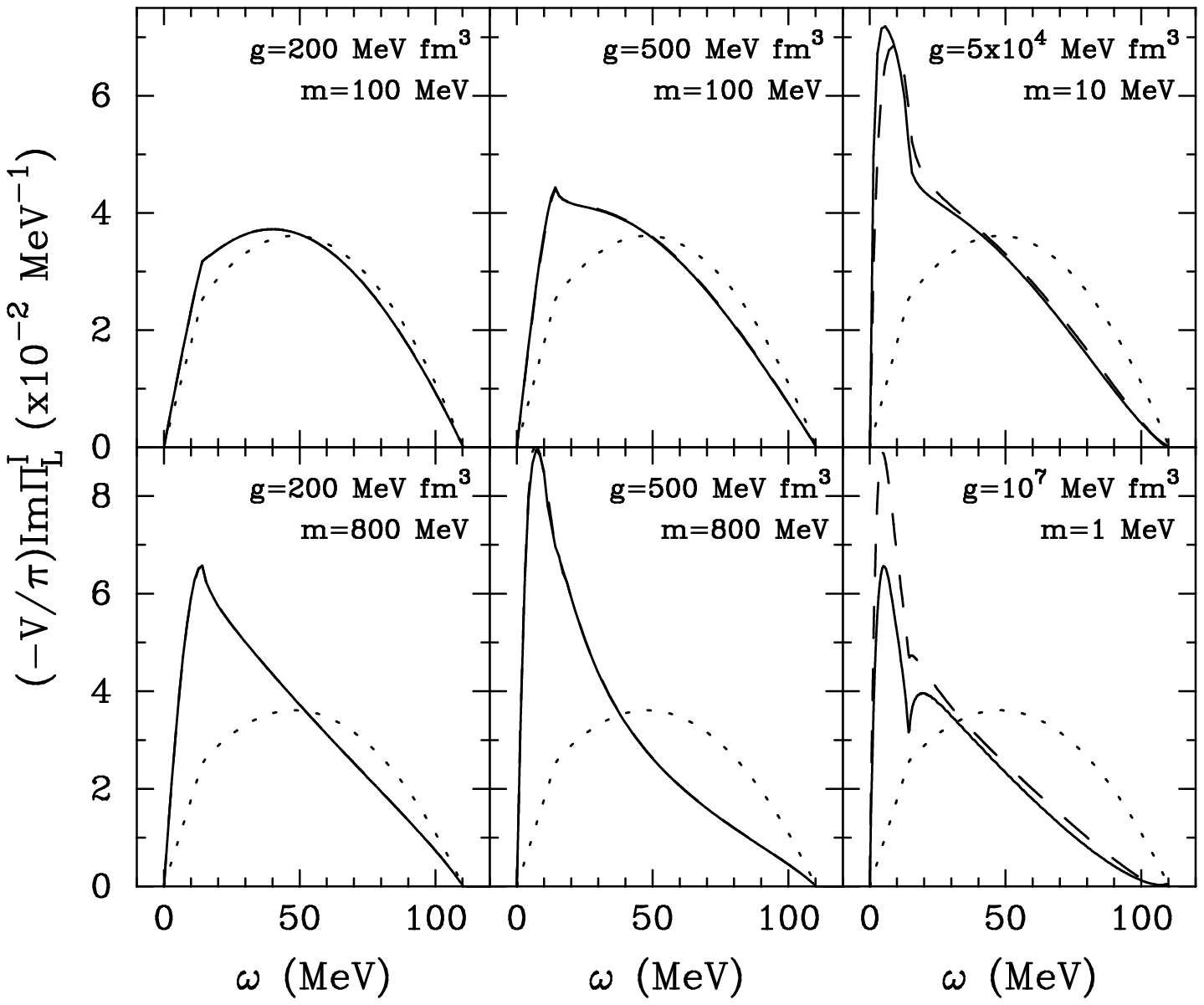,width=.62\textwidth}}
\vskip 2mm
\caption{ Fermi gas longitudinal responses for $k_F=195$ MeV/c at $q=300$ MeV/c,
with a spin-spin one-boson-exchange interaction, for various values of the
coupling constant and of the boson mass: Free response (dot), RPA with the first
order CF expansion (dash) and RPA with the second order CF expansion (solid).
Note that in the left and middle panels the dashed and solid lines are not
distinguishable. The kinematics is non-relativistic.
 }
\label{fig:cf2sigma}
\vfill
\mbox{\epsfig{file=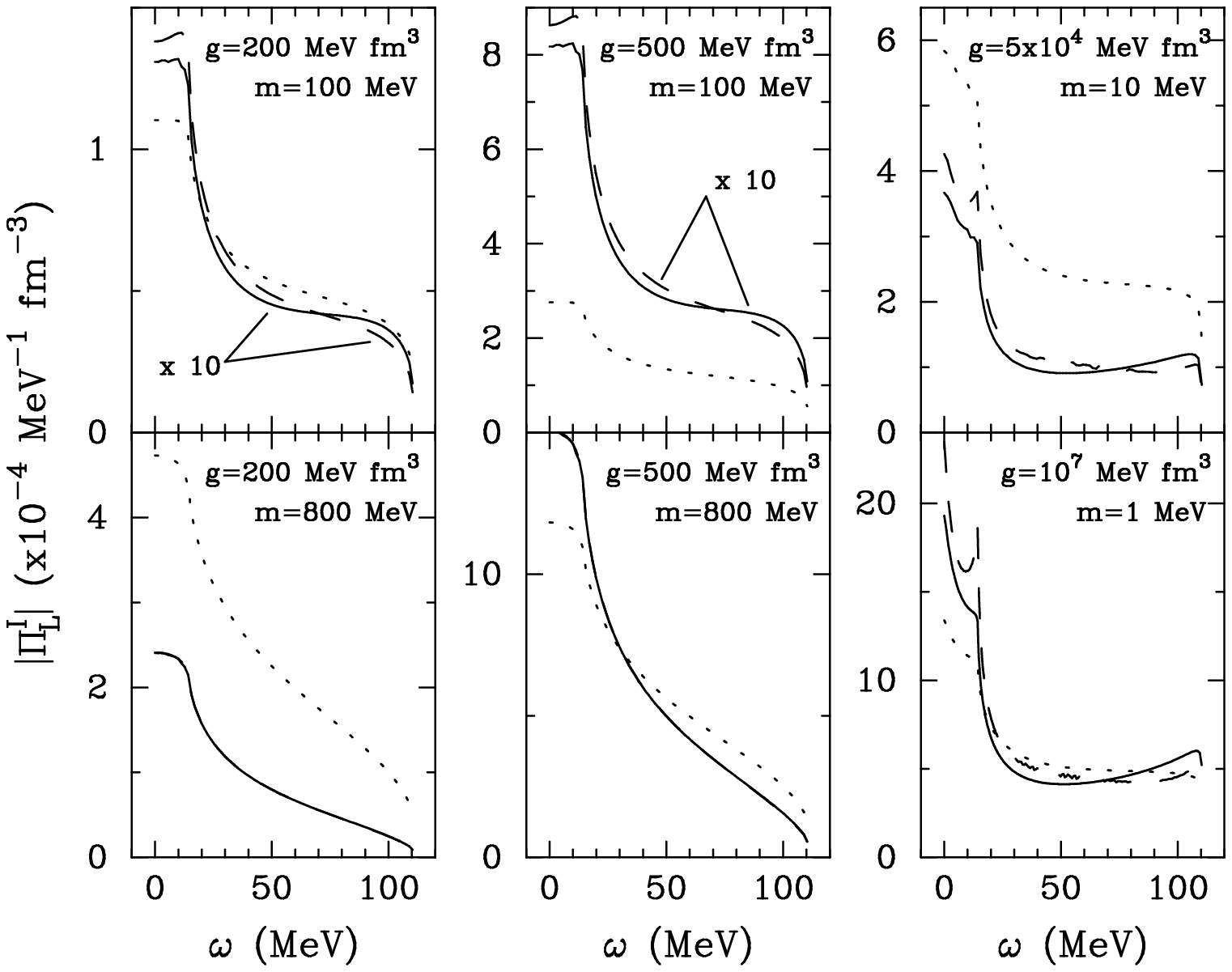,width=.62\textwidth}}
\vskip 2mm
\caption{ Modulus of the longitudinal polarization propagator for $k_F=195$
MeV/c at $q=300$ MeV/c, with a spin-spin one-boson-exchange interaction, for
various values of the coupling constant and of the boson mass: First order,
$\Pi^{\text{(1)}}$ (dot); exact second order, $\Pi^{\text{(2)}}$ (dash);
CF approximation to the second order,
$\Pi^{\text{(2)appr}}={\Pi^{\text{(1)}}}^2/\Pi^{\text{(0)}}$ (solid).
 }
\label{fig:r1vs2sigma}
\end{center}
\end{figure}

Let us start by testing the convergence of the CF expansion. 
For this purpose, we compare in Fig.~\ref{fig:cf2sigma} the
longitudinal RPA responses at first and second order in the CF expansion using a
model one-boson-exchange interaction,
$V_\sigma(k)=\bbox{\sigma}_1\cdot\bbox{\sigma}_2 g [m^2/(m^2+k^2)]$ (the spin
operators having the purpose of killing the direct (ring) contribution).
For values of the coupling constant $g$ and of the boson mass $m$ typical of
realistic nucleon-nucleon potentials one finds that the first and second order 
results match at the level of a few per cent (in the left and middle panels of
Fig.~\ref{fig:cf2sigma}, the solid and dashed curved are actually
indistinguishable). One has to go to very low boson masses (a few MeV) and,
consequently, to very high values of $g$ in order to find some
discrepancies. To understand better these results, we display in 
Fig.~\ref{fig:r1vs2sigma} the modulus of the polarization propagator at first
order, $\Pi^{(1)}$ (dot), at second order, $\Pi^{(2)}$ (dash) and the
approximation to $\Pi^{(2)}$ generated by the first order CF expansion (see
Section \ref{subsec:RPA-resp}), 
$\Pi^{(2)\text{appr}}\equiv{\Pi^{(1)}}^2/\Pi^{(0)}$ (solid). From inspection of
the curves, it is clear that the first important element to guarantee a good
convergence is the range of the interaction: Indeed, for $m=800$ MeV
(short-range) $\Pi^{(2)}$ and $\Pi^{(2)\text{appr}}$ practically coincide
independently of the strength of the interaction. This, of course, should be
expected, since for zero-range interactions the first order CF expansion gives
the exact result. For masses of the order of the pion mass one starts finding
discrepancies between $\Pi^{(2)}$ and $\Pi^{(2)\text{appr}}$: However, for
realistic values of the interaction strength the second order contribution turns
out to be one order of magnitude smaller than the first order one, thus making
these discrepancies having no effect on the full response functions 
(Fig.~\ref{fig:cf2sigma}).

To understand these results  it may be useful to compare the strength of the
interactions employed here to the one of one-pion-exchange, 
$g_\pi m_\pi^2\equiv f_\pi^2/3\cong0.33$ (in natural units). With the same
units, the cases with $m=100$ MeV correspond to $g\,m^2=0.26$ and 0.65; those
with $m=800$ MeV to $g\,m^2=16.7$ and 41.7; for $m=1$ and 10 MeV one has 
$g\,m^2=1.3$ and 0.65, respectively.

To summarize, from the left and middle panels of Fig.~\ref{fig:r1vs2sigma} one
can understand that the validity of the CF expansion originates out of the
interplay between range and strength of the interaction: For short-range
potentials, --- where the conventional perturbative expansion may not converge,
--- the CF technique gives a good approximation of the propagators at all 
orders; for long-range (on the nuclear scale) forces, the CF approximation is 
less accurate, but the relative weakness of the interaction already guarantees 
the convergence of the conventional perturbative expansion. One has to go to 
unreasonably low masses to find a situation where the interaction range is very
long and $\Pi^{(1)}$ and $\Pi^{(2)}$ are of the same order (right panels in 
Fig.~\ref{fig:r1vs2sigma}).

We can thus conclude that accurate calculations of nuclear response functions
in the antisymmetrized RPA can be performed at first order in the CF expansion.
The same conclusion is supported also by calculations with a realistic effective
interaction, --- such as the $G$-matrix parameterization discussed below, ---
and including HF and relativistic kinematical effects.

Finally, it is interesting and important to test the validity of the ring
approximation, --- where exchange diagrams are not included, --- since this
approximation has been widely used in the literature because of its simplicity.
In this scheme, the effect of antisymmetrization is mimicked by adding to the
direct interaction matrix elements an effective exchange contribution (see, 
e.~g., Ref.~\cite{Ose82}). For details see also Ref.~\cite{DeP97}, where a
prescription, suitable for the quasifree region, to determine the effective
exchange momentum has been given.

\begin{figure}[t]
\begin{center}
\mbox{\epsfig{file=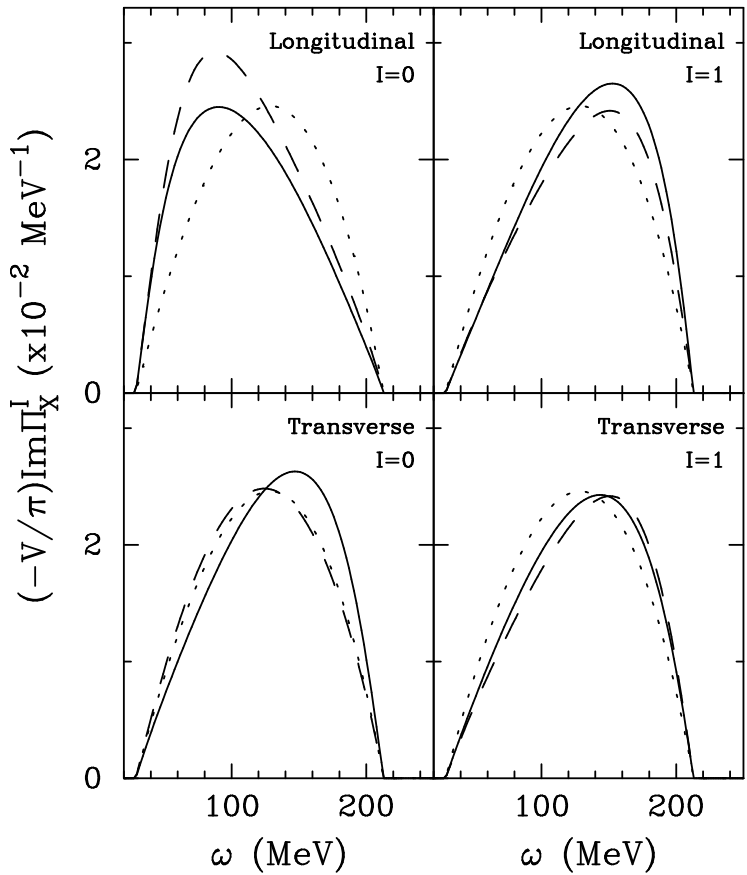,width=.6\textwidth}}
\vskip 2mm
\caption{ Fermi gas responses for $k_F=195$ MeV/c at $q=500$ MeV/c, with the
$G$-matrix parameterization discussed in the text: Free response (dot), ring 
approximation (dash) and RPA (solid). The kinematics is relativistic.
 }
\label{fig:rpavsring}
\end{center}
\end{figure}
In Fig.~\ref{fig:rpavsring}, then, we display the ring and RPA responses of 
$^{12}$C at $q=500$ MeV/c, using the $G$-matrix parameterization. 
It is apparent that the only channel where the ring approximation works 
reasonably well is the spin-isovector one, --- which, incidentally, is the 
dominant one in ($e$,$e'$) magnetic scattering; it is less accurate in all the 
other channels, particularly in the scalar-isoscalar one. The same
considerations apply also when HF correlations are included in the ring and RPA
responses. Note that these results confirm those of 
Ref.~\cite{Shi89}, where a comparison of ring and RPA calculations had been done
using a numerically rather involved finite nucleus formalism. Also in that
calculation the $G$-matrix of Ref.~\cite{Nak84} had been employed.

\section{ Concluding remarks }
\label{sec:concl}

In this paper we have illustrated a fast and compact scheme for the calculation
of the fully antisymmetrized RPA response functions in nuclear matter, based on 
the CF expansion. 
The fast convergence of the CF series for typical nucleon-nucleon potentials has
been demonstrated, thus making this technique a very convenient tool for the
exact resummation of the RPA diagrams. On the other hand the poor performance in
most spin-isospin channels of the ring approximation, --- where the exchange 
diagrams are not included, --- has been confirmed.
Accurate approximations for the inclusion of the relativistic kinematics and of
HF effects have also been discussed and tested. 

Although other classes of contributions are also necessary to make contact with
the electron scattering phenomenology, --- such as meson exchange currents and
higher order ph configurations, --- we believe that the methods discussed in
this paper provide a good, --- because of the accuracy, --- and convenient, ---
because of the simplicity, --- starting point.
As mentioned in the Introduction, it would also be interesting a comparison,
under the same approximation schemes, with other approaches, such as those based
on the relativistic models of nuclear structure.

\appendix

\section{ Tensor interaction in the exchange diagrams }
\label{app:C}

The $n$-th order exchange polarization propagator in presence of tensor
interactions has an expression slightly more complicated than 
(\ref{eq:Pinexpsi}), because the tensor operators do not allow, in general, for
a factorization of the azimuthal integrations.
A generic diagram with $m$ non-tensor and $n-m$ tensor interaction lines can
instead be written as 
\begin{eqnarray}
  \Pi^{(n)\text{ex}}_{\alpha_1...\alpha_m,\alpha_{m+1}...\alpha_n}(q,\omega) &=&
    (-1)^n\left(\frac{m_N}{q}\right)^{n+1}\left(\frac{k_F}{2\pi}\right)^{2n+2}
    \nonumber \\ && \times
    \int_{-1}^{1}dy_1\frac{1}{2}\int_{0}^{1-y_1^2}dx_1\cdot\cdot\cdot
    \int_{-1}^{1}dy_{n+1}\frac{1}{2}\int_{0}^{1-y_{n+1}^2}dx_{n+1} \nonumber \\
  && \times\frac{1}{\psi-y_1+i\eta_{\omega}}W_{\alpha_1}(x_1,y_1;x_2,y_2)
   \cdot\cdot\cdot W_{\alpha_m}(x_m,y_m;x_{m+1},y_{m+1}) \nonumber \\
  && \times W_{\alpha_{m+1}...\alpha_n}(x_{m+1},y_{m+1};...;x_{n+1},y_{n+1})
   \frac{1}{\psi-y_{n+1}+i\eta_{\omega}} \nonumber \\
  && + \sum(\omega\to-\omega) ,
\end{eqnarray}
where $W_{\alpha_i}$ has been defined in (\ref{eq:Walpha}) for the non-tensor
channels and
\begin{eqnarray}
  && W_{\alpha_{m+1}...\alpha_n}(x_{m+1},y_{m+1};...;x_{n+1},y_{n+1}) =
    2^{n-m}\sum_{ij}\sum_{l_1...l_{n-m}}\Lambda_{ji}
    \int_0^{2\pi}\frac{d\varphi_{m+1}}{2\pi}...
    \int_0^{2\pi}\frac{d\varphi_{n+1}}{2\pi} \nonumber \\
  && \qquad\times 
    V_{\alpha_{m+1}}(\bbox{k}_{m+1}-\bbox{k}_{m+2})
      S_{il_1}(\widehat{\bbox{k}_{m+1}-\bbox{k}_{m+2}}) ...
    V_{\alpha_{n}}(\bbox{k}_{n}-\bbox{k}_{n+1})
      S_{l_{n-m}j}(\widehat{\bbox{k}_{n}-\bbox{k}_{n+1}}) .
\end{eqnarray}
In the last expression we have introduced the tensors
\begin{equation}
  S_{ij}(\hat{\bbox{k}}) = 3\hat{\bbox{k}}_i \hat{\bbox{k}}j-\delta_{ij} ,
\end{equation}
such that 
$\sum_{ij}\sigma_i\sigma_j S_{ij}(\hat{\bbox{k}}) = S_{12}(\hat{\bbox{k}})$.

The first order case is rather simple, since one gets again 
(\ref{eq:Pi1ex})--(\ref{eq:Wppalpha}) with
\begin{equation}
  W_{\alpha}(x,y;x',y') = \int_0^{2\pi}\frac{d\varphi}{2\pi}
    V_{\alpha}(\bbox{k}-\bbox{k}') S_{zz}(\widehat{\bbox{k}-\bbox{k}'}) .
\label{eq:WalphaTN}
\end{equation}
At second order, however, one can use Eqs.~(\ref{eq:Pi2ex})--(\ref{eq:GWp}) only
when just one tensor interaction is present.

\section{ First and second order exchange diagrams }
\label{app:D}

We give here the explicit expressions for the first and second order exchange 
diagrams, based on the potential (\ref{eq:pot})--(\ref{eq:mes-exch}).
In (\ref{eq:Pi1ex}) and (\ref{eq:Q1alpha}) we have seen that 
\begin{equation}
  \Pi^{(1)\text{ex}}_{\alpha}(q,\omega) =
    -\left(\frac{m_N}{q}\right)^{2}\frac{k_F^4}{(2\pi)^4}
    \left[{\cal Q}_\alpha^{(1)}(0,\psi)
    - {\cal Q}_\alpha^{(1)}(\bar{q},\psi)
    + {\cal Q}_\alpha^{(1)}(0,\psi+\bar{q})
    - {\cal Q}_\alpha^{(1)}(-\bar{q},\psi+\bar{q})\right]
    ,
\end{equation}
where 
\begin{equation}
  {\cal Q}_\alpha^{(1)}(\bar{q},\psi) = 2 \int_{-1}^1 dy
    \frac{1}{\psi-y+i\eta_{\omega}}\int_{-1}^1 dy' \, {W_\alpha}''(y,y';\bar{q})
    \frac{1}{y-y'+\bar{q}} ,
\end{equation}
whereas from (\ref{eq:Pi2ex}) and (\ref{eq:Q2aap}) one has
\begin{eqnarray}
  && \Pi^{(2)\text{ex}}_{\alpha\alpha'}(q,\omega) =
    \left(\frac{m_N}{q}\right)^{3}\frac{k_F^6}{(2\pi)^6}
    \left[{\cal Q}_{\alpha\alpha'}^{(2)}(0,0;\psi)
        - {\cal Q}_{\alpha\alpha'}^{(2)}(0,\bar{q};\psi)
        - {\cal Q}_{\alpha\alpha'}^{(2)}(\bar{q},0;\psi)
        + {\cal Q}_{\alpha\alpha'}^{(2)}(\bar{q},\bar{q};\psi) \right.
  \nonumber \\
  &&\qquad \left.
        - {\cal Q}_{\alpha\alpha'}^{(2)}(0,0;\psi+\bar{q})
        + {\cal Q}_{\alpha\alpha'}^{(2)}(0,-\bar{q};\psi+\bar{q})
        + {\cal Q}_{\alpha\alpha'}^{(2)}(-\bar{q},0;\psi+\bar{q})
        - {\cal Q}_{\alpha\alpha'}^{(2)}(-\bar{q},-\bar{q};\psi+\bar{q})
    \right] , \nonumber \\
\end{eqnarray}
where 
\begin{equation}
  {\cal Q}_{\alpha\alpha'}^{(2)}(\bar{q}_1,\bar{q}_2;\psi) = \int_{-1}^1 dy
    \frac{1}{2}\int_{0}^{1-y^2} dx \, 
    {\cal G}_{\alpha}(x,y+\bar{q}_1;\psi+\bar{q}_1)
    \frac{1}{\psi-y+i\eta_{\omega}}
    {\cal G}_{\alpha'}(x,y+\bar{q}_2;\psi+\bar{q}_2)
\end{equation}
and
\begin{equation}
  {\cal G}_{\alpha}(x,y;\psi) = \int_{-1}^1 dy'
    \frac{1}{\psi-y'+i\eta_{\omega}}W'_\alpha(x,y;y') .
\end{equation}
For a meson-exchange potential the quantities that can be calculated 
analytically are those given by Eqs.~(\ref{eq:Walpha}), (\ref{eq:WalphaTN}),
(\ref{eq:Wpalpha}) and (\ref{eq:Wppalpha}), namely
\begin{mathletters}
\begin{eqnarray}
  W_{\alpha}(x,y;x',y') &=& \int_0^{2\pi}\frac{d\varphi}{2\pi}
    V_{\alpha}(\bbox{k}-\bbox{k}')\phantom{S_{zz}(\widehat{\bbox{k}-\bbox{k}'})}
    \qquad\text{(non-tensor)} \\
  W_{\alpha}(x,y;x',y') &=& \int_0^{2\pi}\frac{d\varphi}{2\pi}
    V_{\alpha}(\bbox{k}-\bbox{k}') S_{zz}(\widehat{\bbox{k}-\bbox{k}'}) 
    \qquad\text{(tensor)} 
\end{eqnarray}
\end{mathletters}
and
\begin{eqnarray}
  W'_\alpha(x,y;y') &=& \frac{1}{2}\int_0^{1-{y'}^2}dx' \, W_\alpha(x,y;x',y')
    \\
  W''_\alpha(y,y';\bar{q}) &=& \frac{1}{2}\int_0^{1-y^2}dx\,
    \frac{1}{2}\int_0^{1-{y'}^2}dx' \, W_\alpha(x,y+\bar{q};x',y') .
\end{eqnarray}
In any channel $\alpha$ the potential is expressed as a combination of the terms
displayed in (\ref{eq:mes-exch}). Then, for each of them one finds
\begin{mathletters}
\begin{eqnarray}
  W_{\delta}(x,y;x',y') &=& \left\{
    \begin{array}{ll}
      g_{\delta} , 
        & \quad \ell=0 \\
      g_{\delta} (\lambda^2-\mu^2) w_a(\lambda|x,y;x',y') ,
        & \quad \ell=1 \\
      g_{\delta} (\lambda^2-\mu^2)^2 w_b(\lambda|x,y;x',y') , 
        & \quad \ell=2
    \end{array}
    \right. \\
  W_{\text{MD}}(x,y;x',y') &=& \left\{
    \begin{array}{ll}
      g_{\text{MD}}\, \mu^2 w_a(\mu|x,y;x',y') , 
        & \quad \ell=0 \\
      g_{\text{MD}}\, \mu^2 [w_a(\mu|x,y;x',y')-w_a(\lambda|x,y;x',y')] , 
        & \quad \ell=1 \\
      g_{\text{MD}}\, \mu^2 [w_a(\mu|x,y;x',y')-w_a(\lambda|x,y;x',y') & \\
        \quad - (\lambda^2-\mu^2) w_b(\lambda|x,y;x',y')] , 
        & \quad \ell=2
    \end{array}
    \right. \\
  W_{\text{TN}}(x,y;x',y') &=& \left\{
    \begin{array}{ll}
      g_{\text{TN}} \{[3(y-y')^2+\mu^2] w_a(\mu|x,y;x',y') - 1\} , 
        & \quad \ell=0 \\
      g_{\text{TN}} \{[3(y-y')^2+\mu^2] w_a(\mu|x,y;x',y') & \\
        \quad - [3(y-y')^2+\lambda^2] w_a(\lambda|x,y;x',y')\} , 
        & \quad \ell=1 \\
      g_{\text{TN}} \{[3(y-y')^2+\mu^2] & \\
        \quad\times [w_a(\mu|x,y;x',y') - w_a(\lambda|x,y;x',y')] & \\
        \quad - (\lambda^2-\mu^2) 
        [3(y-y')^2+\lambda^2] w_b(\lambda|x,y;x',y')\} , 
        & \quad \ell=2 ,
    \end{array}
    \right. 
\end{eqnarray}
\end{mathletters}
where again $\ell$ labels the power of the form factors, we have introduced 
the adimensional form factor cut-off, $\lambda=\Lambda/k_F$, and meson mass, 
$\mu=m/k_F$, and we have defined 
\begin{mathletters}
\begin{eqnarray}
  w_a(\lambda|x,y;x',y') &=&
    \{ [ \lambda^2+(y-y')^2 +x+x' ]^2 - 4xx' \}^{-1/2}
    \\
  w_b(\lambda|x,y;x',y') &=&
    \frac{\lambda^2+(y-y')^2 +x+x'}
         { \{ [ \lambda^2+(y-y')^2 +x+x' ]^2 - 4xx' \}^{3/2} }
    .
\end{eqnarray}
\end{mathletters}

For $W'_{\alpha}$ one finds
\begin{mathletters}
\begin{eqnarray}
  W'_{\delta}(x,y;y') &=& \left\{
    \begin{array}{ll}
      g_{\delta} (1-{y'}^2)/2, 
        & \quad \ell=0 \\
      g_{\delta} (\lambda^2-\mu^2) w'_a(\lambda|x,y;y') ,
        & \quad \ell=1 \\
      g_{\delta} (\lambda^2-\mu^2)^2 w'_b(\lambda|x,y;y') , 
        & \quad \ell=2
    \end{array}
    \right. \\
  W'_{\text{MD}}(x,y;y') &=& \left\{
    \begin{array}{ll}
      g_{\text{MD}}\, \mu^2 w'_a(\mu|x,y;y') , 
        & \quad \ell=0 \\
      g_{\text{MD}}\, \mu^2 [w'_a(\mu|x,y;y')-w'_a(\lambda|x,y;y')] , 
        & \quad \ell=1 \\
      g_{\text{MD}}\, \mu^2 [w'_a(\mu|x,y;y')-w'_a(\lambda|x,y;y') & \\
        \quad - (\lambda^2-\mu^2) w'_b(\lambda|x,y;y')] , 
        & \quad \ell=2
    \end{array}
    \right. \\
  W'_{\text{TN}}(x,y;y') &=& \left\{
    \begin{array}{ll}
      g_{\text{TN}} \{[3(y-y')^2+\mu^2] w'_a(\mu|x,y;y') - (1-{y'}^2)/2\} , 
        & \quad \ell=0 \\
      g_{\text{TN}} \{[3(y-y')^2+\mu^2] w'_a(\mu|x,y;y') & \\
        \quad - [3(y-y')^2+\lambda^2] w'_a(\lambda|x,y;y')\} , 
        & \quad \ell=1 \\
      g_{\text{TN}} \{[3(y-y')^2+\mu^2] [w'_a(\mu|x,y;y')
        - w'_a(\lambda|x,y;y')] & \\
        \quad - (\lambda^2-\mu^2) 
        [3(y-y')^2+\lambda^2] w'_b(\lambda|x,y;y')\} , 
        & \quad \ell=2 ,
    \end{array}
    \right. 
\end{eqnarray}
\end{mathletters}
where
\begin{mathletters}
\begin{eqnarray}
  && w'_a(\lambda|x,y;y') = \frac{1}{2} \nonumber \\
  && \quad \times \ln\left|
    \frac{ \lambda^2+(y-y')^2 +1-{y'}^2-x+
    \sqrt{ [ \lambda^2+(y-y')^2 +1-{y'}^2+x ]^2 - 4(1-{y'}^2)x } }
    {2 [ \lambda^2+(y-y')^2 ] } \right| 
    \nonumber \\
  \\
  && w'_b(\lambda|x,y;y') = \frac{1}{4}\frac{1}{\lambda^2+(y-y')^2}
    \left[1-\frac{\lambda^2+(y-y')^2-1+{y'}^2+x }
    { \sqrt{ [ \lambda^2+(y-y')^2 +1-{y'}^2+x ]^2 - 4(1-{y'}^2)x } } \right]
    .
    \nonumber \\
\end{eqnarray}
\end{mathletters}

Finally, for $W''_{\alpha}$ one finds
\begin{mathletters}
\begin{eqnarray}
  W''_{\delta}(y,y';\bar{q}) &=& \left\{
    \begin{array}{ll}
      g_{\delta} [(1-y^2)/2] [(1-{y'}^2)/2], 
        & \quad \ell=0 \\
      g_{\delta} (\lambda^2-\mu^2) w''_a(\lambda|y,y';\bar{q}) ,
        & \quad \ell=1 \\
      g_{\delta} (\lambda^2-\mu^2)^2 w''_b(\lambda|y,y';\bar{q}) , 
        & \quad \ell=2
    \end{array}
    \right. \\
  W''_{\text{MD}}(y,y';\bar{q}) &=& \left\{
    \begin{array}{ll}
      g_{\text{MD}}\, \mu^2 w''_a(\mu|y,y';\bar{q}) , 
        & \quad \ell=0 \\
      g_{\text{MD}}\, \mu^2 
          [w''_a(\mu|y,y';\bar{q})-w''_a(\lambda|y,y';\bar{q})] , 
        & \quad \ell=1 \\
      g_{\text{MD}}\, \mu^2 [w''_a(\mu|y,y';\bar{q})-w''_a(\lambda|y,y';\bar{q})
        - (\lambda^2-\mu^2) w''_b(\lambda|y,y';\bar{q})] , 
        & \quad \ell=2
    \end{array}
    \right. \nonumber \\ \\
  W''_{\text{TN}}(y,y';\bar{q}) &=& \left\{
    \begin{array}{ll}
      g_{\text{TN}} \{[3(y-y'+\bar{q})^2+\mu^2] w''_a(\mu|y,y';\bar{q}) & \\
        \quad -[(1-y^2)/2] [(1-{y'}^2)/2] \} , 
        & \quad \ell=0 \\
      g_{\text{TN}} \{[3(y-y'+\bar{q})^2+\mu^2] w''_a(\mu|y,y';\bar{q}) & \\
        \quad - [3(y-y'+\bar{q})^2+\lambda^2] w''_a(\lambda|y,y';\bar{q})\} , 
        & \quad \ell=1 \\
      g_{\text{TN}} \{[3(y-y'+\bar{q})^2+\mu^2] [w''_a(\mu|y,y';\bar{q})
        - w''_a(\lambda|y,y';\bar{q})] & \\
        \quad - (\lambda^2-\mu^2) 
        [3(y-y'+\bar{q})^2+\lambda^2] w''_b(\lambda|y,y';\bar{q})\} , 
        & \quad \ell=2 ,
    \end{array}
    \right. 
\end{eqnarray}
\end{mathletters}
where
\begin{mathletters}
\begin{eqnarray}
  && w''_a(\lambda|y,y';\bar{q}) = \frac{1}{8} \Big\{ \nonumber \\
  &&-[ \lambda^2+(y-y'+\bar{q})^2 +2-y^2-{y'}^2 ]
    -2(2-y^2-{y'}^2)\ln|2[\lambda^2+(y-y'+\bar{q})^2]| \nonumber \\
  &&+\sqrt{[ \lambda^2+(y-y'+\bar{q})^2 +2-y^2-{y'}^2 ]^2-4(1-y^2)(1-{y'}^2)} 
    \nonumber \\
  &&+2(1-y^2) \nonumber \\
  && \times\ln\left|\lambda^2+(y-y'+\bar{q})^2+y^2-{y'}^2
    +\sqrt{[ \lambda^2+(y-y'+\bar{q})^2 +2-y^2-{y'}^2 ]^2-4(1-y^2)(1-{y'}^2)} 
    \right| \nonumber \\
  &&+2(1-{y'}^2) \nonumber \\
  && \times\ln\left|\lambda^2+(y-y'+\bar{q})^2-y^2+{y'}^2
    +\sqrt{[ \lambda^2+(y-y'+\bar{q})^2 +2-y^2-{y'}^2 ]^2-4(1-y^2)(1-{y'}^2)} 
    \right| \Big\} \nonumber \\
    \\
  && w''_b(\lambda|y,y';\bar{q}) = \frac{1}{8} \nonumber \\
  && \times \frac{ \lambda^2+(y-y'+\bar{q})^2 +2-y^2-{y'}^2 
    -\sqrt{[ \lambda^2+(y-y'+\bar{q})^2 +2-y^2-{y'}^2 ]^2-4(1-y^2)(1-{y'}^2)} }
    {\lambda^2+(y-y'+\bar{q})^2}
    .
    \nonumber \\
\end{eqnarray}
\end{mathletters}


\begin{references}

\bibitem{Jou96}    J. Jourdan,
                   Nucl.\ Phys.\ {\bf A603} (1996) 117.

\bibitem{Ang96}    M. Anghinolfi {\em et al\/.},
                   Nucl.\ Phys.\ {\bf A602} (1996) 405.

\bibitem{Wil97}    C. F. Williamson {\em et al\/.},
                   Phys.\ Rev.\ C {\bf 56} (1997) 3152.

\bibitem{Del85}    A. Dellafiore, F. Lenz, and F. A. Brieva, 
                   Phys.\ Rev.\ {\bf C31} (1985) 1088.

\bibitem{Del87}    F. A. Brieva and A. Dellafiore, 
                   Phys.\ Rev.\ {\bf C36} (1987) 899.

\bibitem{Shi89}    T. Shigehara, K. Shimizu, and A. Arima, 
                   Nucl.\ Phys.\ {\bf A492} (1989) 388.

\bibitem{Hor90}    C. J. Horowitz and J. Piekarewicz,
                   Nucl.\ Phys.\ {\bf A511} (1990) 461.

\bibitem{Bub91}    M. Buballa, S. Drozdz, S. Krewald, and J. Speth,
                   Ann.\ Phys.\ (N.Y.) {\bf 208} (1991) 346.

\bibitem{Weh93}    K. Wehrberger,
                   Phys.\ Rep. {\bf 225} (1993) 273.

\bibitem{Ama94}    J. E. Amaro, G. C\`o, and A. M. Lallena,
                   Nucl.\ Phys.\ {\bf A578} (1994) 365.

\bibitem{Cai95}    J. C. Caillon and J. Labarsouque,
                   Nucl.\ Phys.\ {\bf A595} (1995) 189.

\bibitem{ Kim95}   Hungchong Kim, C. J. Horowitz, and M. R. Frank,
                   Phys.\ Rev.\ {\bf C51} (1995) 792.

\bibitem{Bar96a}   M. B. Barbaro, A. De Pace, T. W. Donnelly,
                   and A. Molinari, 
                   Nucl.\ Phys.\ {\bf A596} (1996) 553.

\bibitem{Bes96}    J. Besprosvany,
                   Nucl.\ Phys.\ {\bf A601} (1996) 269.

\bibitem{Fab97}    A. Fabrocini,
                   Phys.\ Rev.\ {\bf C55} (1997) 338.

\bibitem{Cen97}    R. Cenni, F. Conte, and P. Saracco,
                   Nucl.\ Phys.\ {\bf A623} (1997) 391.

\bibitem{Gil97}    A. Gil, J. Nieves, and E. Oset,
                   Nucl.\ Phys.\ {\bf A627} (1997) 599.

\bibitem{Fet71}    A. L. Fetter and J. D. Walecka,
                   {\em Quantum Theory of Many-Particle Systems}
                   (McGraw-Hill, New York, 1971).

\bibitem{Wal95}    J. D. Walecka,
                   {\em Theoretical Nuclear and Subnuclear Physics}
                   (Oxford University Press, Oxford, 1995).

\bibitem{Bar96b}   M. B. Barbaro, A. De Pace, T. W. Donnelly,
                   and A. Molinari, 
                   Nucl.\ Phys.\ {\bf A598} (1996) 503.

\bibitem{Amo96}    P. Amore, M. B. Barbaro, and A. De Pace
                   Phys.\ Rev.\ C {\bf 53} (1996) 2801.

\bibitem{Nak84}    K. Nakayama, S. Krewald, J. Speth, and W. G. Love,
                   Nucl.\ Phys.\ {\bf A431} (1984) 419.

\bibitem{Nak87}    K. Nakayama, S. Drozdz, S. Krewald, and J. Speth,
                   Nucl.\ Phys.\ {\bf A470} (1987) 573.

\bibitem{Alb93}    W. M. Alberico, M. B. Barbaro, A. De Pace, T. W. Donnelly,
                   and A. Molinari, 
                   Nucl.\ Phys.\ {\bf A563} (1993) 605.

\bibitem{Bar94}    M. B. Barbaro, A. De Pace, T.W. Donnelly, and A. Molinari, 
                   Nucl.\ Phys.\ {\bf A569} (1994) 701.
                                                                
\bibitem{Gal58}    V. M. Galitskii and A. B. Migdal, 
                   Sov.\ Phys.\ JEPT {\bf 34} (1958) 7.

\bibitem{Abr63}    A. A. Abrikosov, L. P. Gorkov, and I. E. Dzyaloshinski, 
                   {\em Methods of Quantum Field Theory in Statistical Physics}
                   (Dover, New York, 1963).

\bibitem{Alb91}    W. M. Alberico, A. De Pace, A. Drago, and A. Molinari,
                   Rivista Nuovo Cimento {\bf 14} (1991) 1.

\bibitem{Alb90}    W. M. Alberico, T. W. Donnelly, and A. Molinari,
                   Nucl.\ Phys.\ {\bf A512} (1990) 541.

\bibitem{Ose82}    E. Oset, H. Toki, and W. Weise, 
                   Phys.\ Rep. {\bf 83} (1982) 281.

\bibitem{Len80}    F. Lenz, E. J. Moniz, and K. Yazaki, 
                   Ann.\ Phys.\ (N.Y.) {\bf 129} (1980) 84.

\bibitem{Fes92}    H. Feshbach, 
                   {\em Theoretical Nuclear Physics: Nuclear Reactions} 
                   (Wiley, New York, 1992).

\bibitem{Bal63}    R. Balescu,
                   {\em Statistical Mechanics of Charged Particles}
                   (Interscience, New York, 1963) p. 399;
                   N. I. Muskhelishvili,
                   {\em Singular Integral Equations}
                   (Noordhoff, Groningen, 1953) pp. 56--61;
                   G. D. White, K. T. R. Davies, and P. J. Siemens,
                   Ann.\ Physics {\bf 187} (1988) 198.

\bibitem{DeP97}    A. De Pace, C. Garc\'\i a-Recio, and E. Oset,
                   Phys.\ Rev.\ {\bf C55} (1997) 1394.

\end{references}
\end{document}